\documentclass{cernyrep}
\usepackage{amsmath}
\usepackage[T1]{fontenc}
\usepackage[bookmarks, colorlinks=true, linktoc=page, pdftex, linkcolor=black, citecolor=black, urlcolor=blue]{hyperref}

\usepackage{fancyhdr}
\fancyhfoffset{4 mm}
\fancypagestyle{ARTTITLE}{% 
\fancyhf{} % clear all header and footer fields
\lhead{Proceedings of the CERN--Accelerator--School course on
{\it Introduction to Accelerator Physics}}
\lfoot{Available online at \url{https://cas.web.cern.ch/previous-schools}}
\rfoot{\thepage\hspace*{3mm}}
 
}
\newcommand{\der}[2]{\frac{\mathrm{d}#1}{\mathrm{d}#2}}

\usepackage{varwidth}
\usepackage{xcolor}

\usepackage[export]{adjustbox}

\begin{document}

\title{Injection and Extraction}
\author{F. Tecker}
\institute{CERN, Geneva, Switzerland}

\begin{abstract}
This paper gives an overview of the beam injection and extraction principles for accelerators. 
After a brief general introduction, it explains different methods of injecting the beam for hadron and lepton machines. 
It describes single- and multi-turn hadron injection, charge-exchange H$^-$ injection, then betatron and synchrotron injection for leptons.
For extraction, it presents single- and multi-turn extraction, as well as resonant extraction methods.
Finally, the requirements for linking several accelerators by a transfer line are presented. 
\end{abstract}
\maketitle
\thispagestyle{ARTTITLE}

\section{Introduction}
Today's accelerators are often very complex machines which are a series of different accelerators 
where the beam is transferred from one to another. Typically, you inject from a linear accelerator into
a circular accelerator, or you transfer the beam from one circular machine into another one in order to
accumulate a bigger beam intensity and/or
accelerate the beam to higher energies. The reason for this is for example that each accelerator has a limited
dynamic range due to magnetic field limitations or that you want to periodically fill a storage ring or a
synchrotron light source. Finally, you also need to extract your beam in a controlled way to a physics experiment
or a beam dump.
So a beam transfer in an out of accelerators is necessary, in which you also have to make sure to preserve
the beam quality and avoid beam loss. 
Many different injection and extraction schemes exist, depending on the particle types and purpose of the transfer,
whether to preserve the transverse and time structure or to modify it. The following sections will give
an overview of the most common schemes. Much more detailed descriptions can be found in a dedicated
CAS course on this topic~\cite{bib:CAS-Inj}.

\section{Injection methods}
When injecting the beam in a circular machine, the main goal is to put the injected beam on the closed orbit of the ring.
Also the shape of the transverse distribution has to match the beam optics of the receiving machine. 
This will be dealt with in section~\ref{sec:linking}. The longitudinal aspect of the injection is covered in~\cite{bib:longitudinal}.

\begin{figure}[tb!]
\begin{center}
\includegraphics[width=13cm]{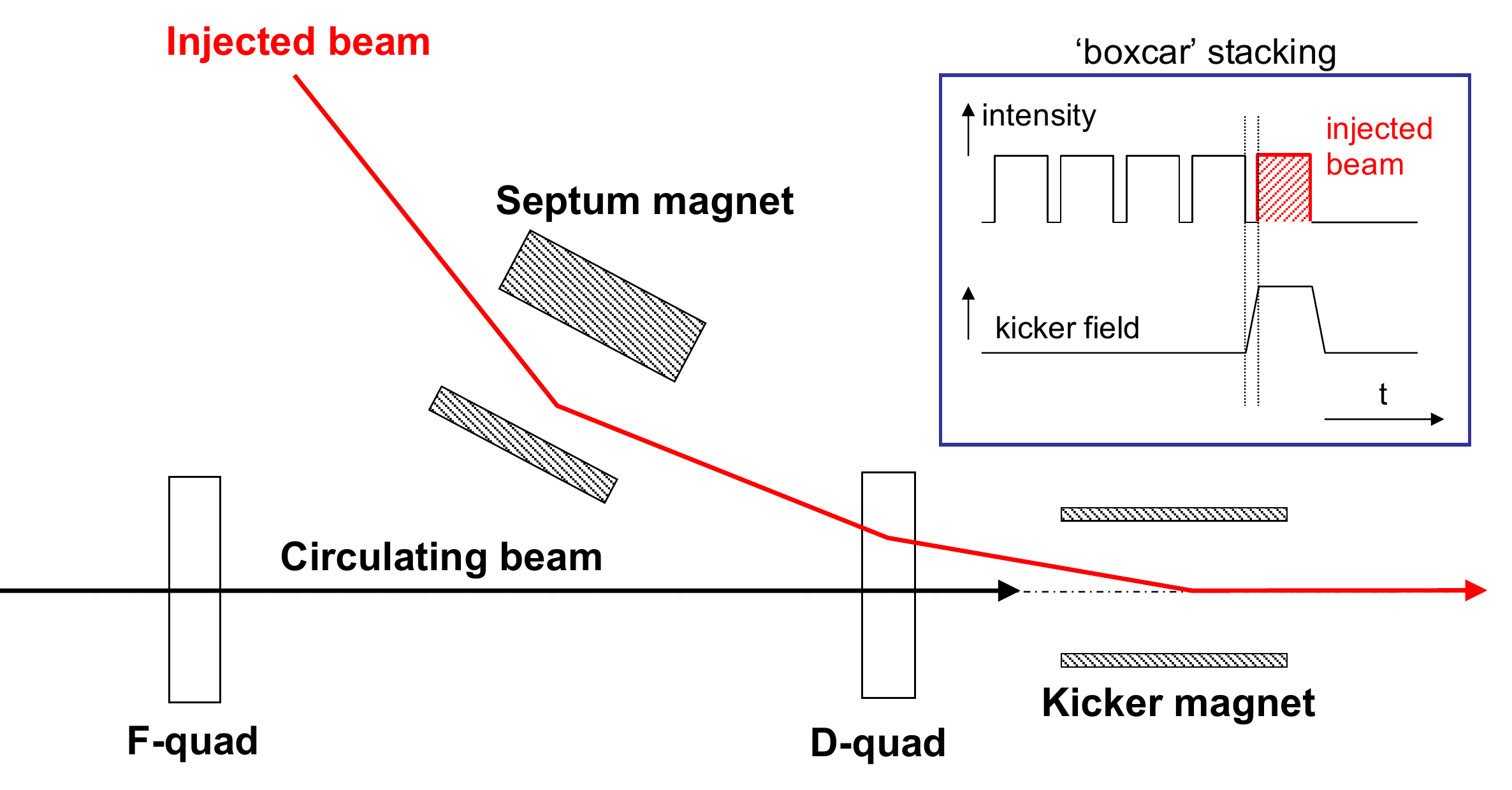}
\end{center}
\caption{
\label{fig:inj-basic}
Basic injection scheme}
\end{figure}

\begin{figure}[htbp!]
\begin{center}
\hfill\begin{minipage}{6cm}
\includegraphics[width=6cm]{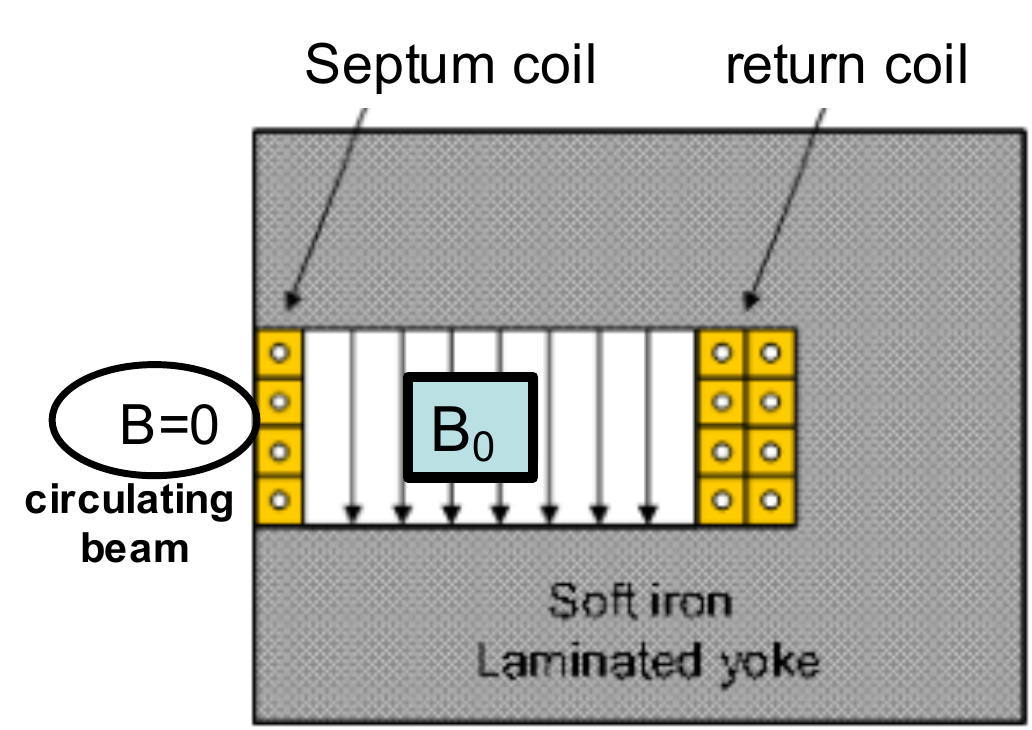}
\end{minipage}\hfill
\begin{minipage}{8cm}
\includegraphics[width=8cm]{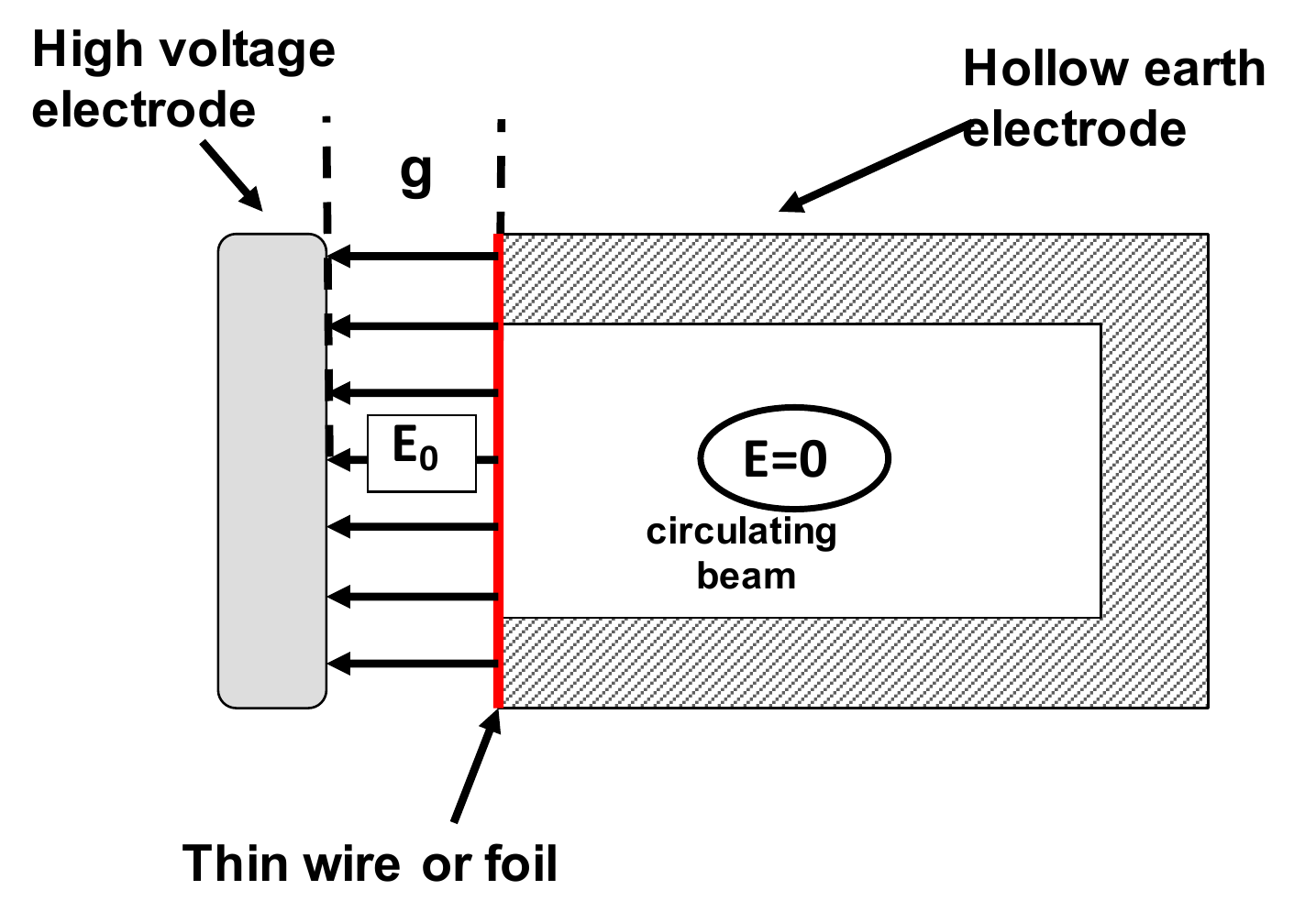}
\end{minipage}\hfill
\end{center}
\caption{
\label{fig:septa}
Schematic view for a magnetic septum (left) and an electrostatic septum (right)}
\end{figure}

The basic scheme of the injection is shown in Fig.~\ref{fig:inj-basic}. 
The beam needs to be brought close to the path of the circulating beam in order to minimise the 
deflection angle required to put it on the closed orbit.
This is why typically a 'septum' is used to deflect the injected beam, where the part separating 
the injected from the circulated beam is very thin. 
A septum can be a magnetic or electrostatic element. Fig.~\ref{fig:septa} shows the basic layout. 
More detailed information can be found in~\cite{bib:septa}.
\

\begin{figure}[htbp!]
\begin{center}
\includegraphics[width=8cm]{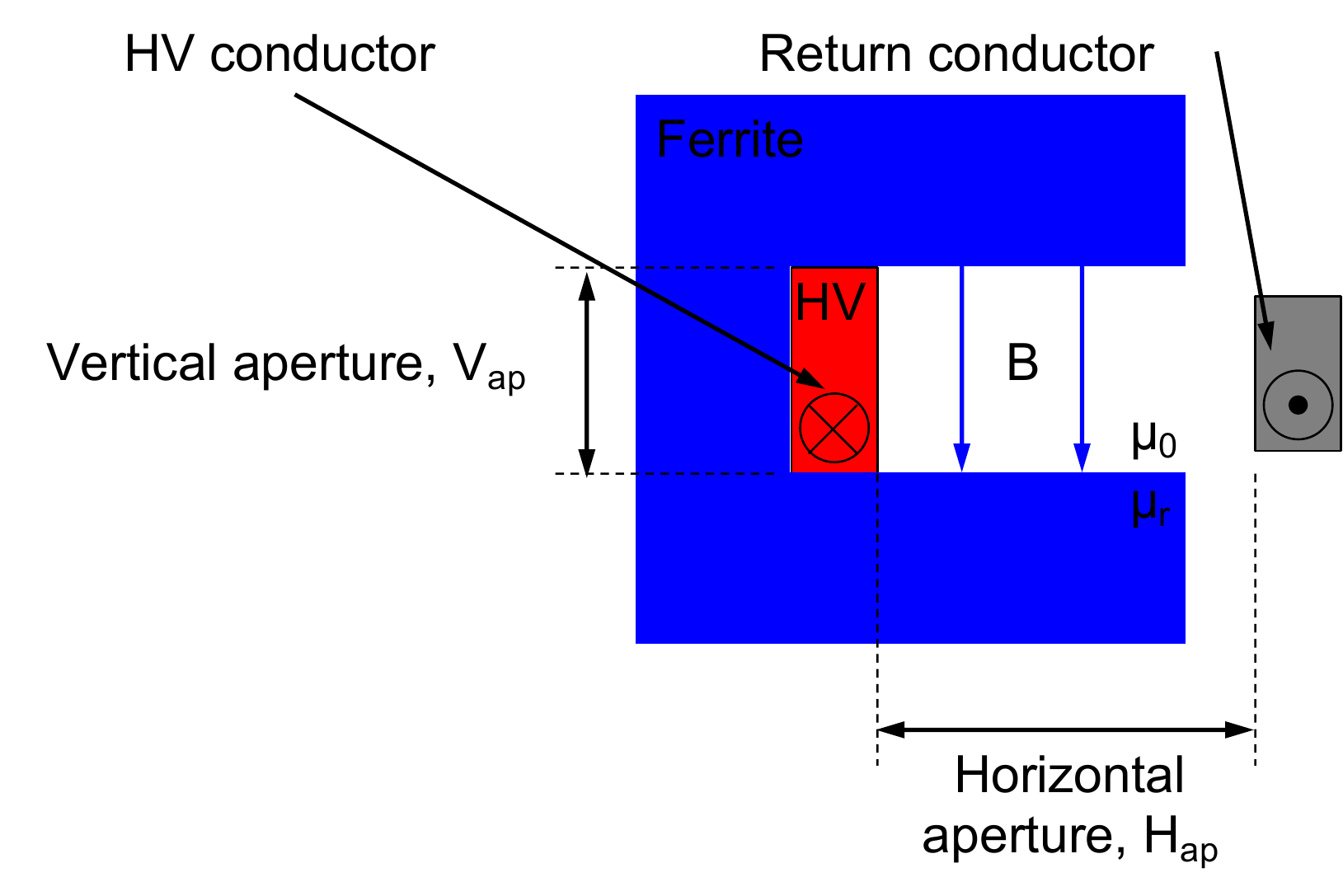}
\end{center}
\caption{
\label{fig:kicker}
Schematic cross-section for a C-core geometry kicker magnet}
\end{figure}

When the injected beam reaches the closed orbit, it needs to be deflected to be put on the proper circulating path.
This is done by a 'kicker', a fast pulsed magnet with a very short rise time (from a few ns - few $\upmu$s).
Fig.~\ref{fig:kicker} shows a schematic view of a kicker magnet.
Ferrite material (permeability $\mu_r \approx 1000$) reinforces the magnetic circuit and the field uniformity
in the gap.
For fast rise-times the inductance must be minimised, and often only one turn of coil is used.
In order to reach the desired total deflection, kickers are often split into several magnet units,
powered independently. More details about kickers can be found in~\cite{bib:kicker}.

%\begin{equation}
%B_y \simeq \mu_0 \left( \frac{N I}{V_\mathrm{ap}} \right)
%\end{equation}
%\begin{equation}
%L_{\rm mag/m} \approx \mu_0 \left( \frac{N^2 H_{\rm ap}}{V_{\rm ap}} \right)
%\end{equation}

\subsection{Single-turn hadron injection}
The single-turn injection is the simplest of the injection schemes. It follows the scheme shown in Fig.~\ref{fig:inj-basic}.
The septum deflects the incoming beam onto the closed orbit at the centre of the kicker,
and the kicker compensates for the remaining angle. Typically, septum and kicker are on either side of
a defocusing quadrupole magnet to minimise the required kicker strength.
The betatron phase advance between septum exit and kicker in the plane of the deflection (more common 
is the horizontal plane) has to be $90^\degree$ for a beam that is parallel to the closed orbit at the septum exit.
In addition, a set of three or four dipole magnets can be used to create a closed orbit bump to approach the
closed orbit to the septum in order to reduce the kicker strength.

The pulse length of the kicker is such that the whole incoming beam pulse is deflected, while the field of the
kicker is not present when the circulating beam passes, so that this one rests unperturbed.
In this way, several beam pulses from the injecting accelerator can be accumulated in the ring longitudinally.
The transverse emittance of the beam is ideally preserved, the total intensity can be increased by the 
successive injections.
 
\begin{figure}[!b]
\centering\includegraphics[width=0.85\linewidth]{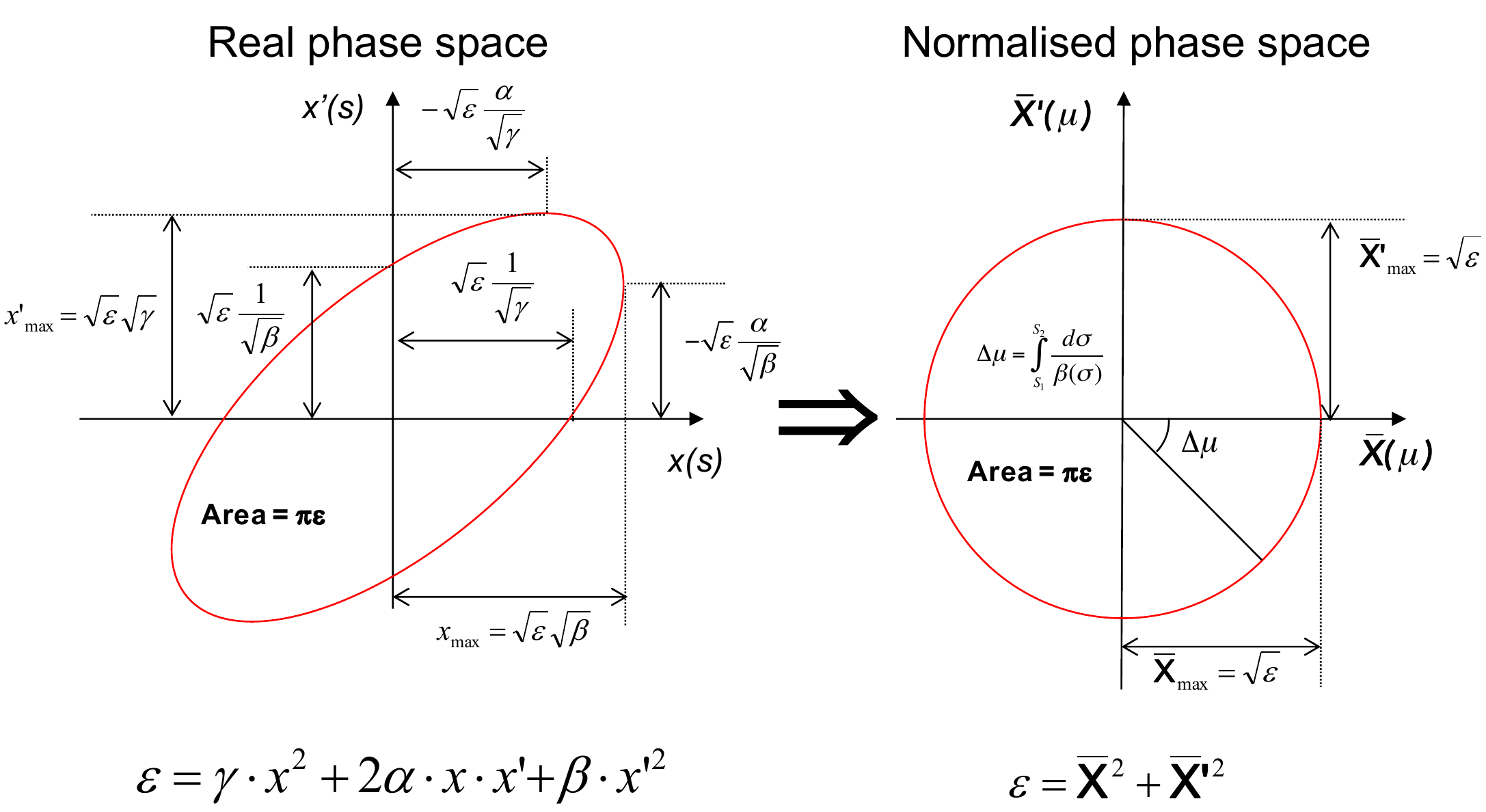}
\caption{Transformation from real to normalised phase space.}
\label{fig:transform}
\end{figure}

In order to describe the injection, it is useful to use normalised phase space coordinates. In this normalization,
the phase space contours, which are generally ellipses, transform into circles all along the accelerator.
This normalisation is given by
\begin{equation}
\begin{pmatrix}
\bar X \\
\bar X'
\end{pmatrix} = \sqrt{\frac{1}{\beta(s)}} 
\begin{pmatrix}
1 & 0 \\
\alpha(s) & \beta(s)
\end{pmatrix}\cdot
\begin{pmatrix}
x \\
x'
\end{pmatrix},
\end{equation}
where $\alpha$ and $\beta$ are the Twiss parameters. The betatron phase advance $\mu$ becomes the independent
variable for the normalised variables:
\begin{eqnarray}
\bar X & = & \sqrt{\frac{1}{\beta(s)}} x = \sqrt{\varepsilon} \cos(\mu - \mu_0) \\
\bar X' & = & \sqrt{\frac{1}{\beta(s)}} \alpha(s) x + \sqrt{\beta(s)} x' = - \sqrt{\varepsilon} \sin(\mu - \mu_0)
= \der{\bar X}{\mu}.
\end{eqnarray}
Fig.~\ref{fig:transform} depicts this transformation, and
\begin{figure}[!tb]
\centering
a)\includegraphics[width=0.33\linewidth]{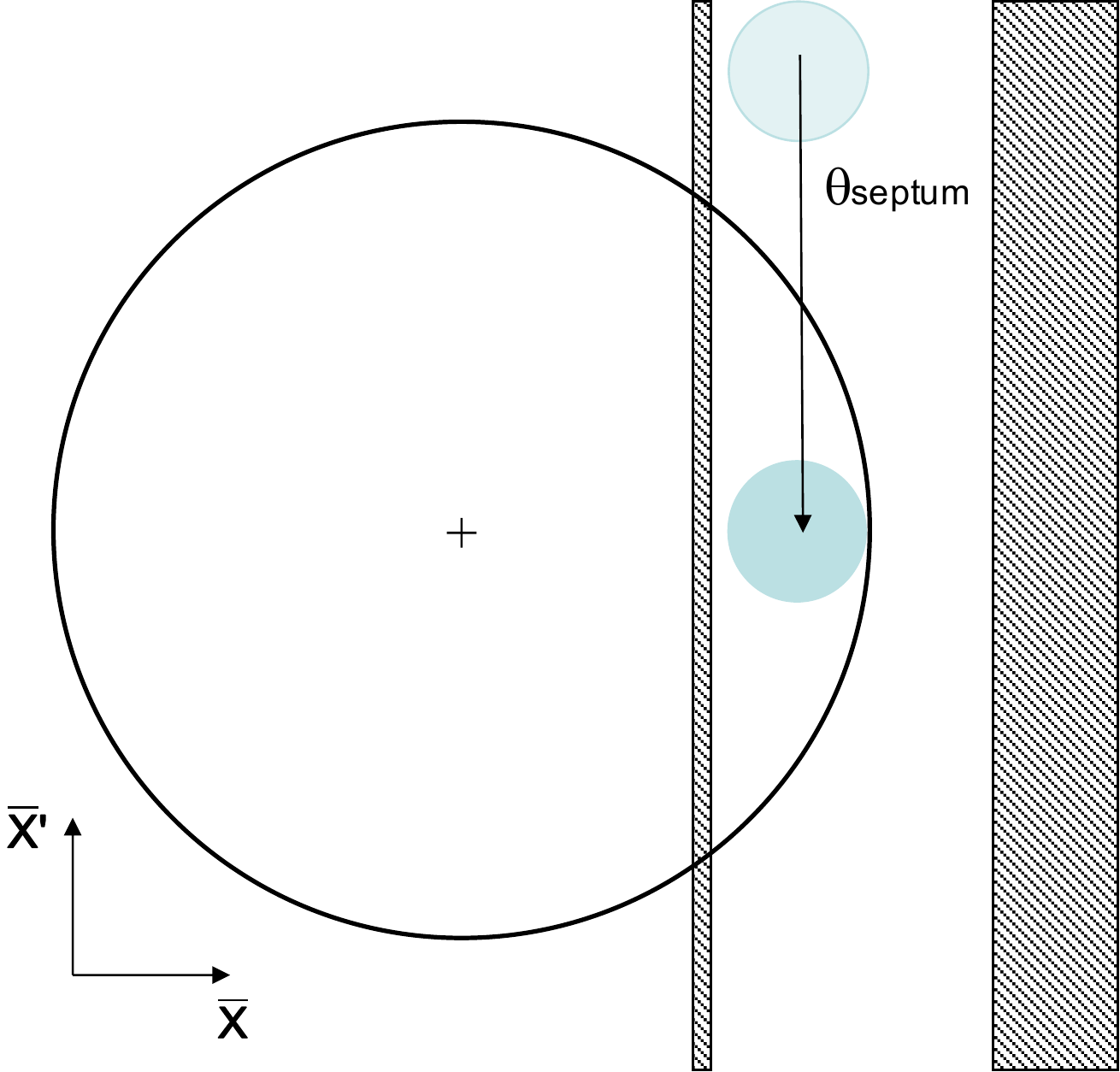}\hfill
b)\includegraphics[width=0.26\linewidth]{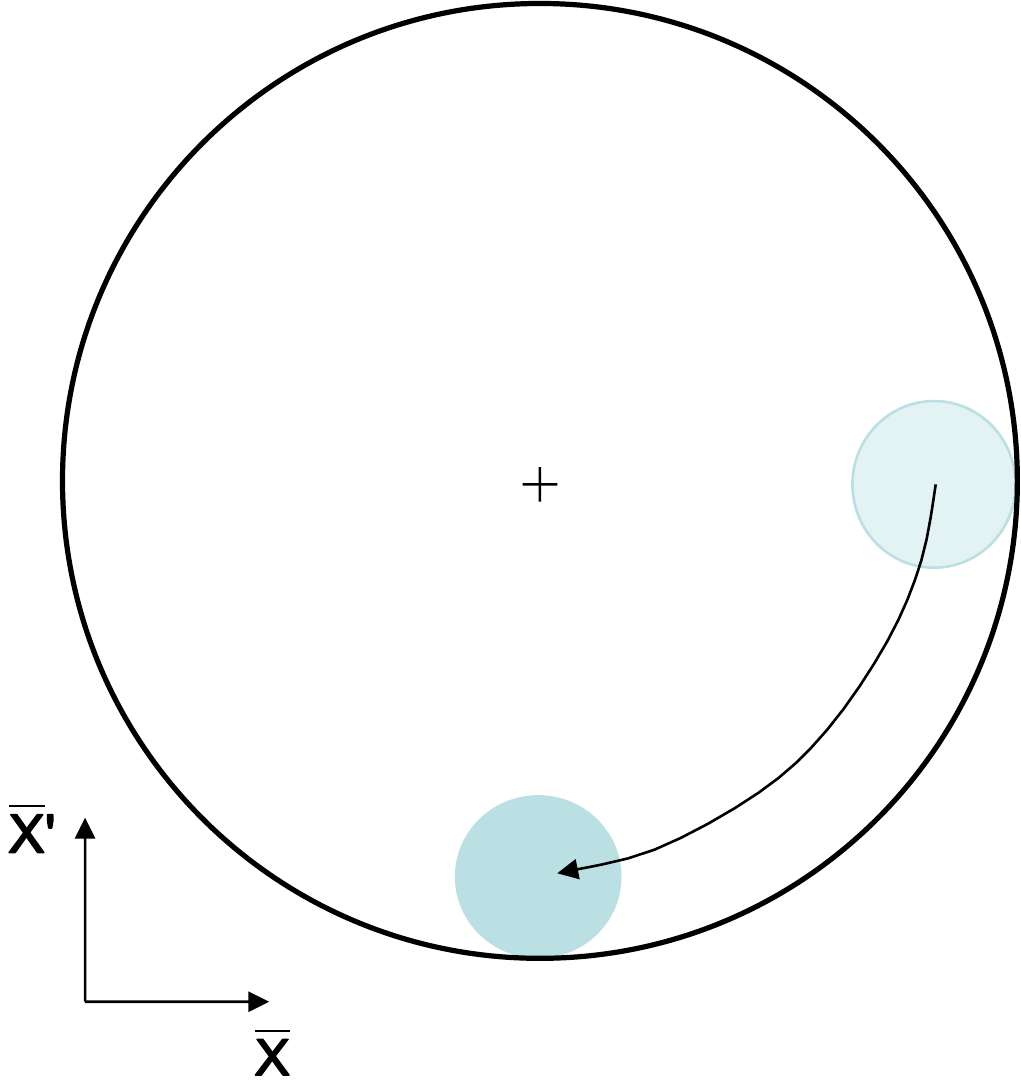}\hfill
c)\includegraphics[width=0.26\linewidth]{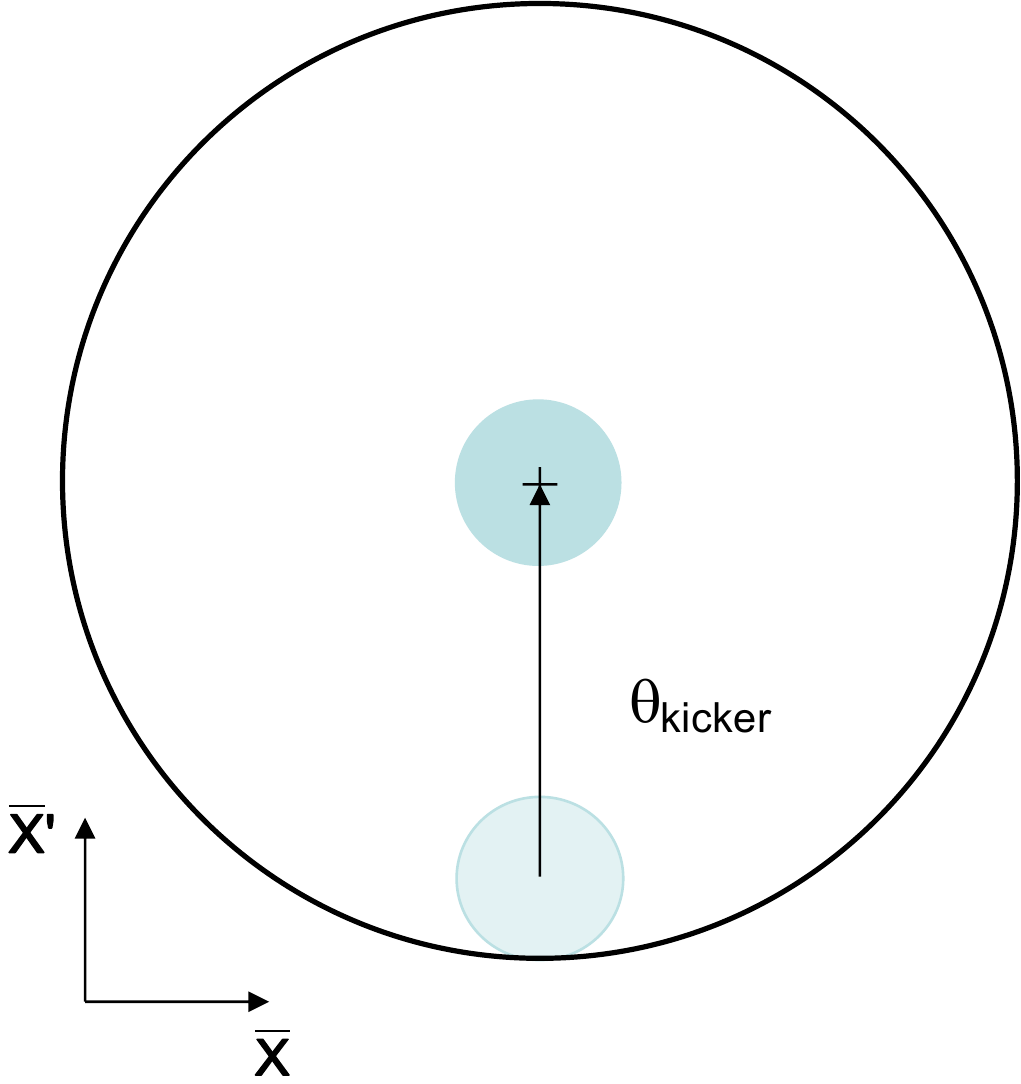}
\caption{Injection process in normalised phase space. a) The septum deflects the incoming beam (blue circle) parallel
to the closed orbit. b) The $90^\degree$ phase advance puts the beam on the closed orbit position.
c) The kicker deflects the beam onto the closed orbit.}
\label{fig:injection-phase-space}
\end{figure}%
the injection process in this normalised phase space is shown in Fig.~\ref{fig:injection-phase-space}.

\subsection{Injection errors, filamentation and blow-up}
It is important for the injection that the beam gets injected onto the closed orbit of the receiving accelerator with
the correct position and correct angle.
Any difference will lead to a betatron oscillation of the beam around this closed orbit. As practically any accelerator has some
higher-order field components, there will be an intrinsic spread in the betatron tune for the particles, 
depending on their oscillation amplitude.
This decoherence of the betatron oscillation results in a larger beam size and effective emittance. 

\begin{figure}[!b]
\centering
a)\includegraphics[width=0.3\linewidth]{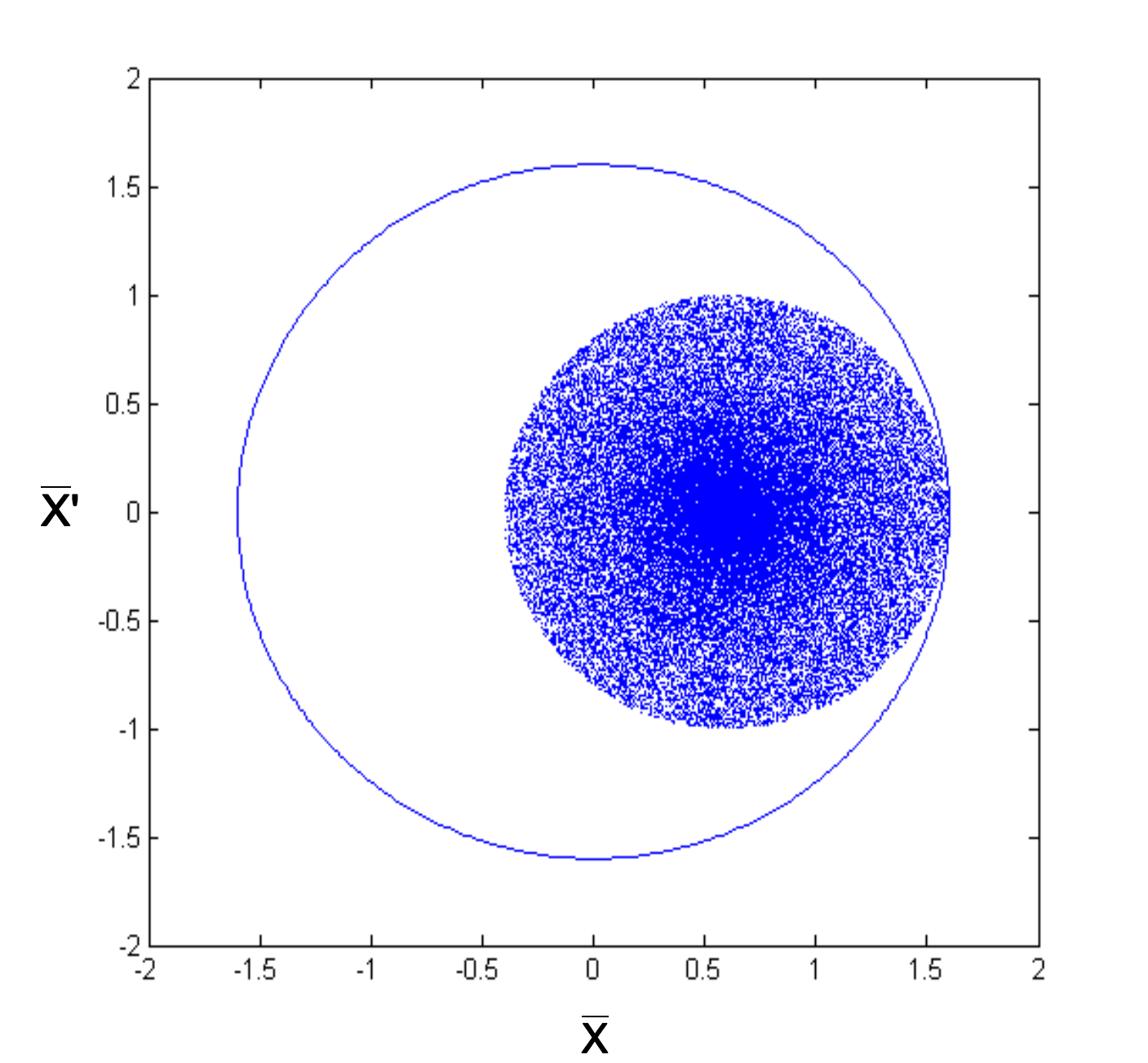}\hfill
b)\includegraphics[width=0.3\linewidth]{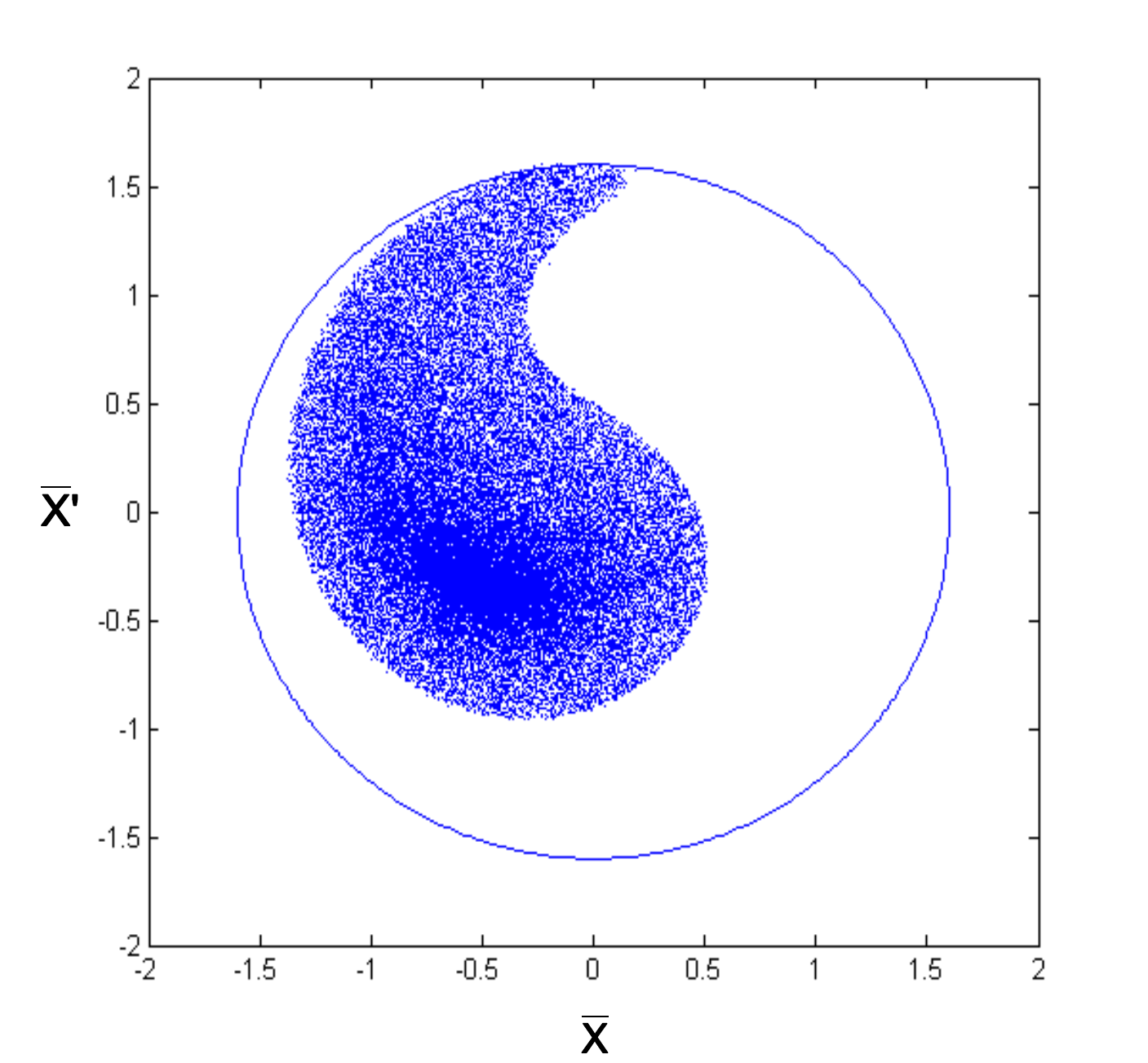}\hfill
c)\includegraphics[width=0.3\linewidth]{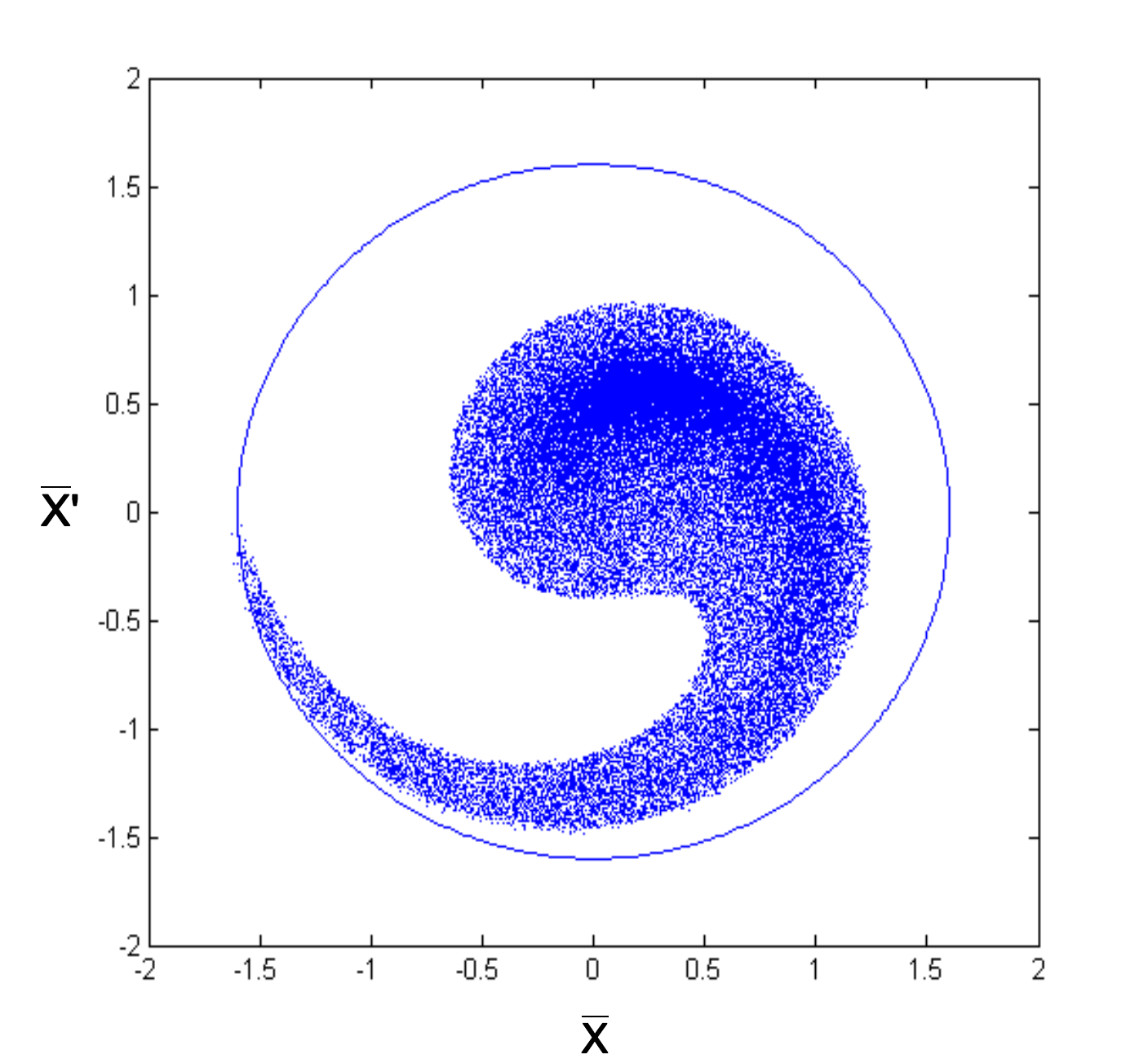}
d)\includegraphics[width=0.3\linewidth]{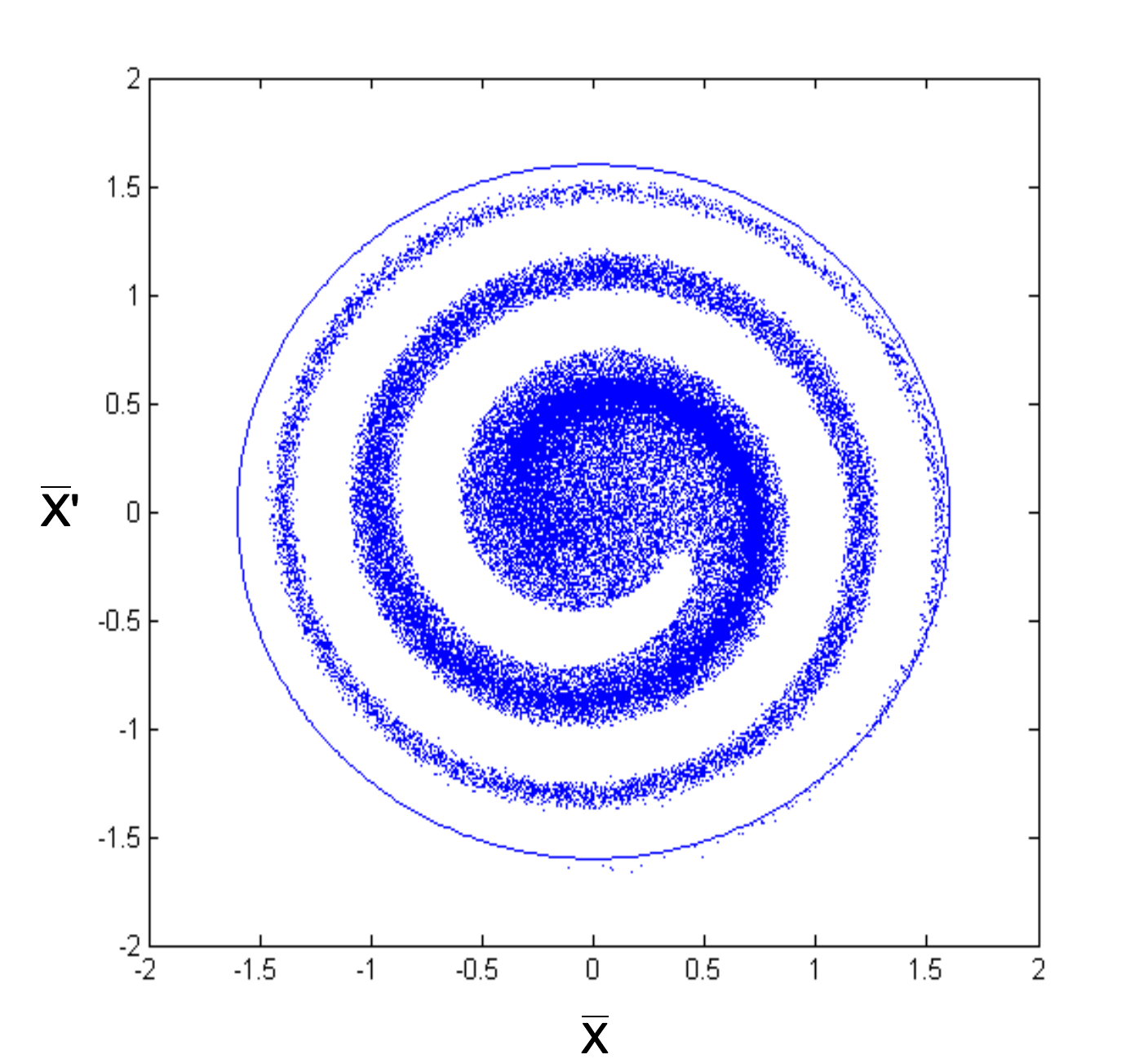}\hfill
e)\includegraphics[width=0.3\linewidth]{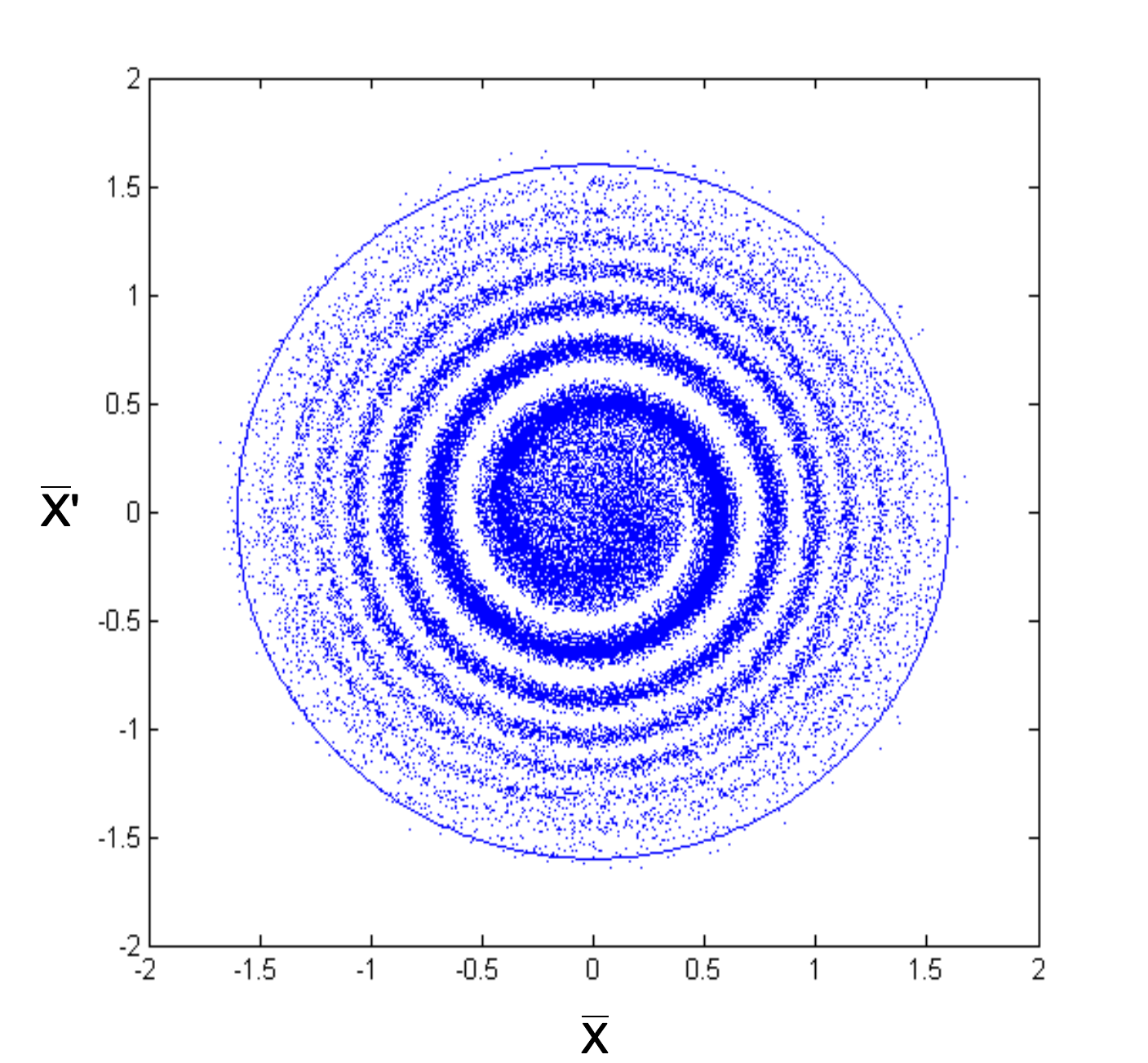}\hfill
f)\includegraphics[width=0.3\linewidth]{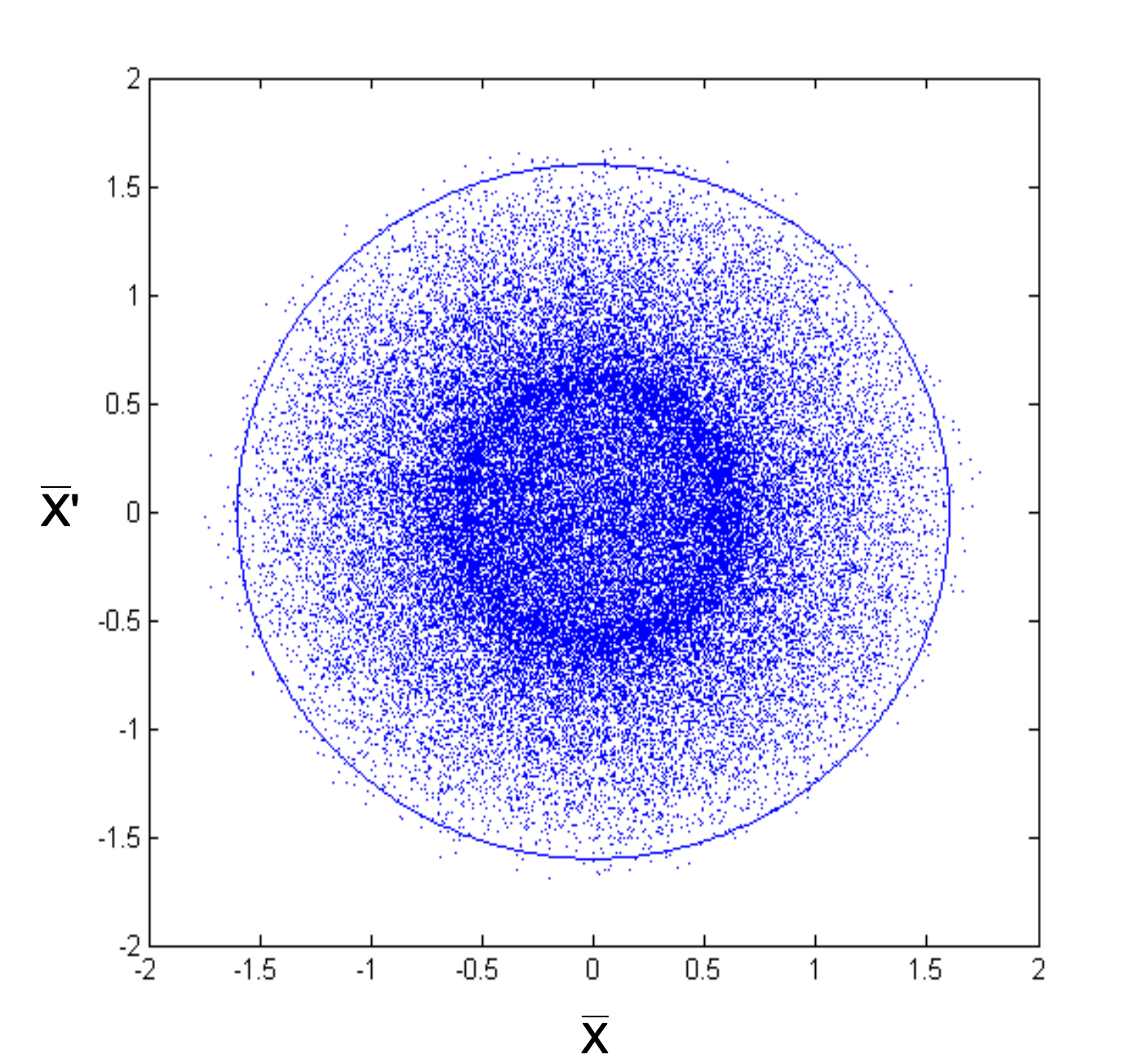}
\caption{Injection with position error in normalised phase space. a) The beam is injection with an error in position.
 b) Due to non-linearities, the phase advance is different for different initial conditions and the distribution
 spreads out after a turn.
 c) The spreading continues after another turn.
 d-e) Over many turns, the phase space contour will further spread and the particles are distributed in 
 an overall larger phase space area.
 f) Finally, particles are filling a much larger phase space area than the initial distribution. }
\label{fig:filamentation}
\end{figure}

The effect is best visualised in phase space (see Fig.~\ref{fig:filamentation}).
The phase advance depends on
the initial condition due to non-linearities. So the phase space distribution of the injected beam gets distorted
if it is not on the closed orbit. This distortion over many turns leads to 'filamentation' of the beam, the beam spreads
out over a larger phase space area, resulting in an increased effective emittance.
 
\begin{figure}[bh!]
\begin{center}
\includegraphics[width=11cm]{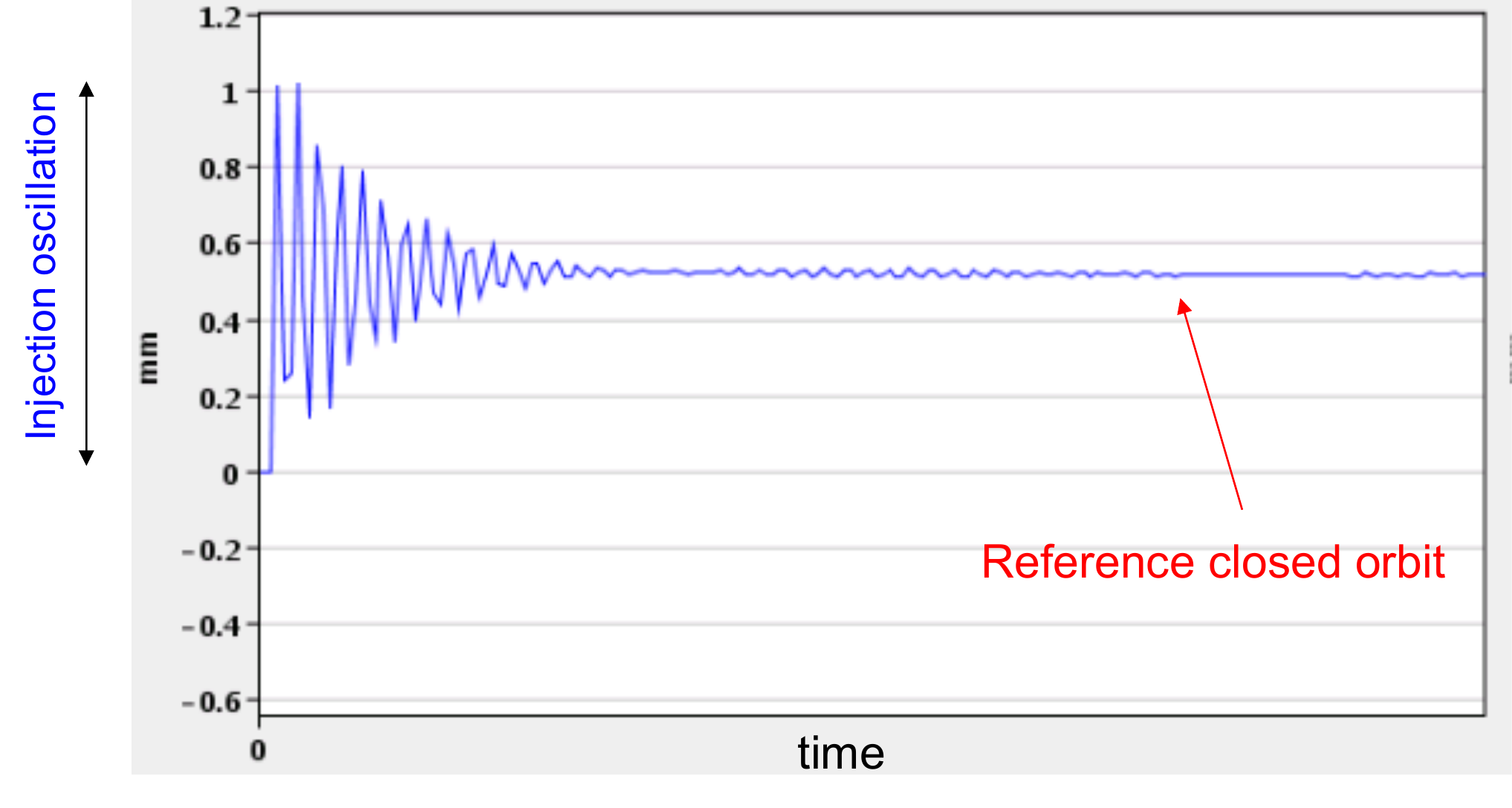}
\end{center}
\caption{Typical time evolution of an injection oscillation measured on a beam position monitor (BPM) in case of
trajectory errors. The oscillation is clearly visible initially, but will become invisible on the BPM, as the individual
oscillations decohere due to a tune spread.
\label{fig:filamentation-orbit}
}
\end{figure}

When these orbit oscillations are observed with a turn-by-turn beam position monitor (BPM) system,
it could give you the impression that the oscillation is damped (see Fig.~\ref{fig:filamentation-orbit}). 
Nevertheless, since the BPMs measure just the centre of gravity of the transverse beam distribution, 
this effect is explained by the decoherence and filamentation, since the individual particles oscillate with 
different frequencies. The decoherence is faster for a larger tune spread, i.e. larger non-linearities
and a larger chromaticity.

The injection can be optimised by observing the oscillation of the injected beam on the BPM system,
either with respect to the closed orbit, or the difference between the first and second turn trajectories.
Since angular errors from septum and kicker show a different orbit oscillation pattern (remember the
$90^\degree$ phase advance between them), the correction can be analytically calculated from the
difference orbit when you have an optics model of your accelerator.

In order to mitigate residual injection errors, a transverse feedback (TFB) or 'transverse damper' system is often used in hadron machines.
This measures the injection oscillation turn-by-turn and applies a kick to the beam to damp the oscillation before the
beam filaments.

\subsection{Multi-turn hadron injection}
When the beam density at injection is limited either by the injector capacity or by space charge effects,
the multi-turn injection process allows to accumulate more bunches in the same bucket.  
It is not possible to inject into the same phase space area, as we would kick out the beam already present there.
But if the acceptance of the receiving machine is larger than the delivered beam emittance, we can fill the
phase space from the inside to the outside by successive injections over several turns.
Multi-turn injection is often performed in the horizontal plane due to the usually larger acceptance.

\begin{figure}[!htbp]
\centering\includegraphics[width=0.85\linewidth]{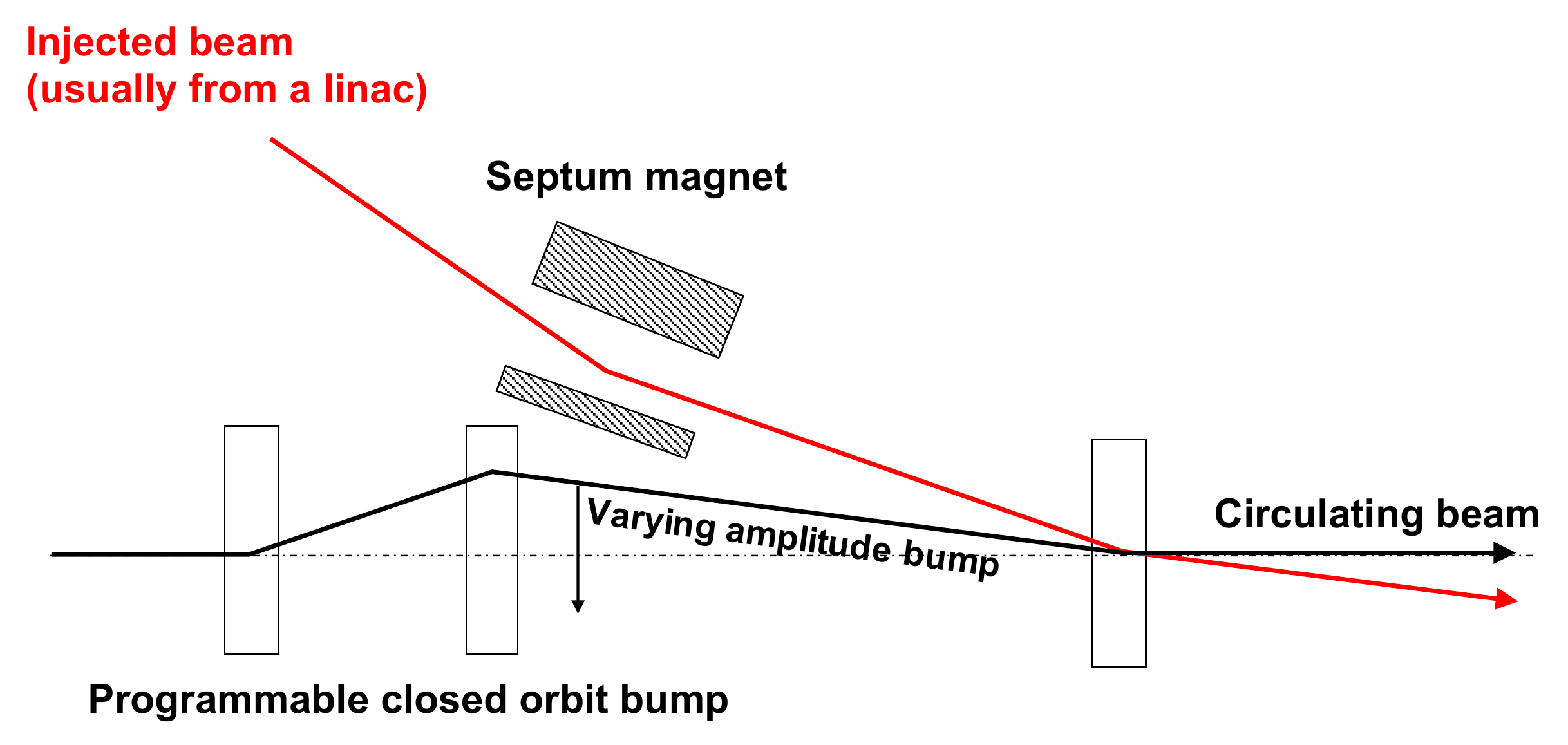}
\caption{Schematic view of the injection region for multi-turn injection.}
\label{fig:multiturn}
\end{figure}

\begin{figure}[!htbp]
\centering
\newcommand{\myfigscale}{0.42}
\includegraphics[width=\myfigscale\linewidth]{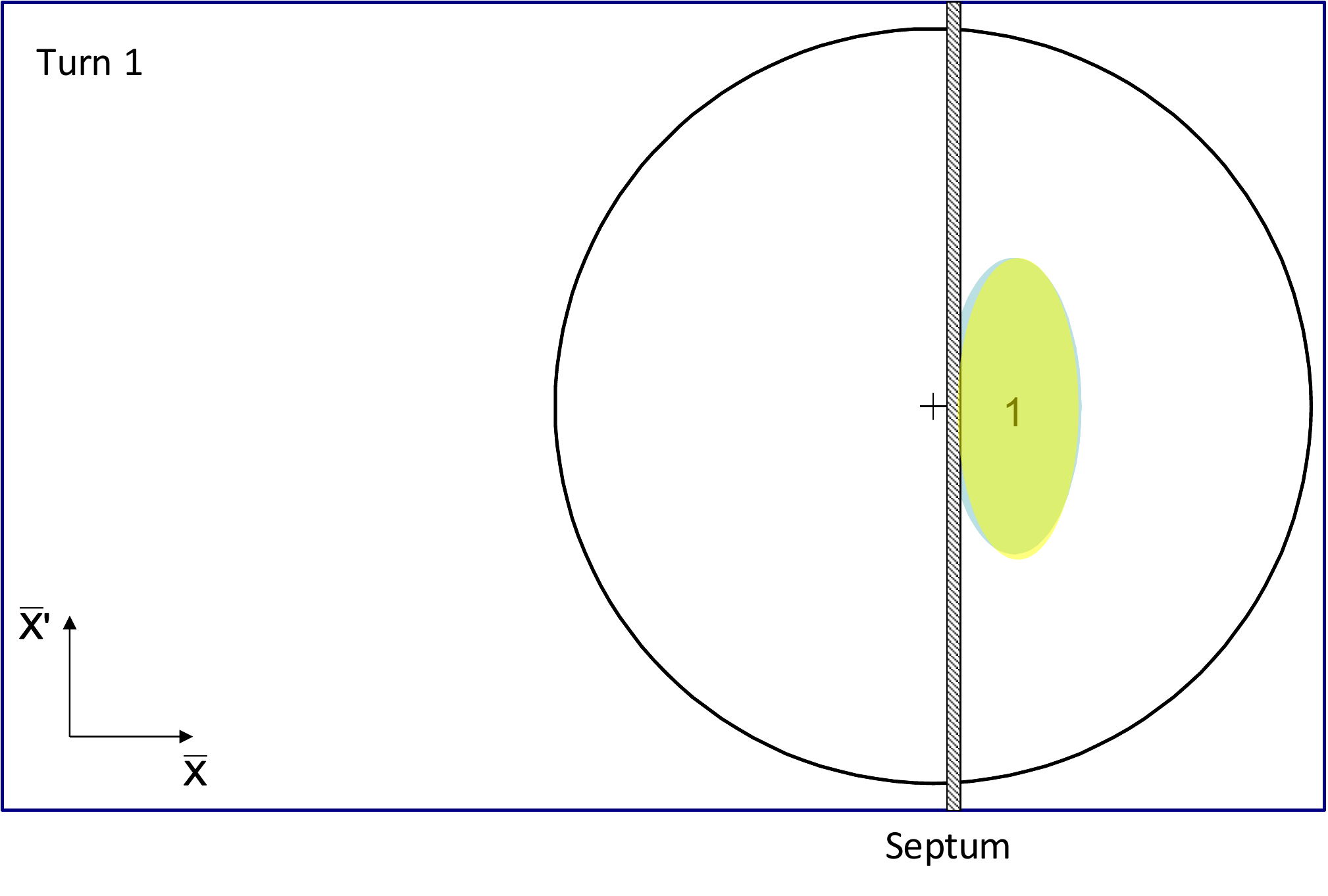}
\includegraphics[width=\myfigscale\linewidth]{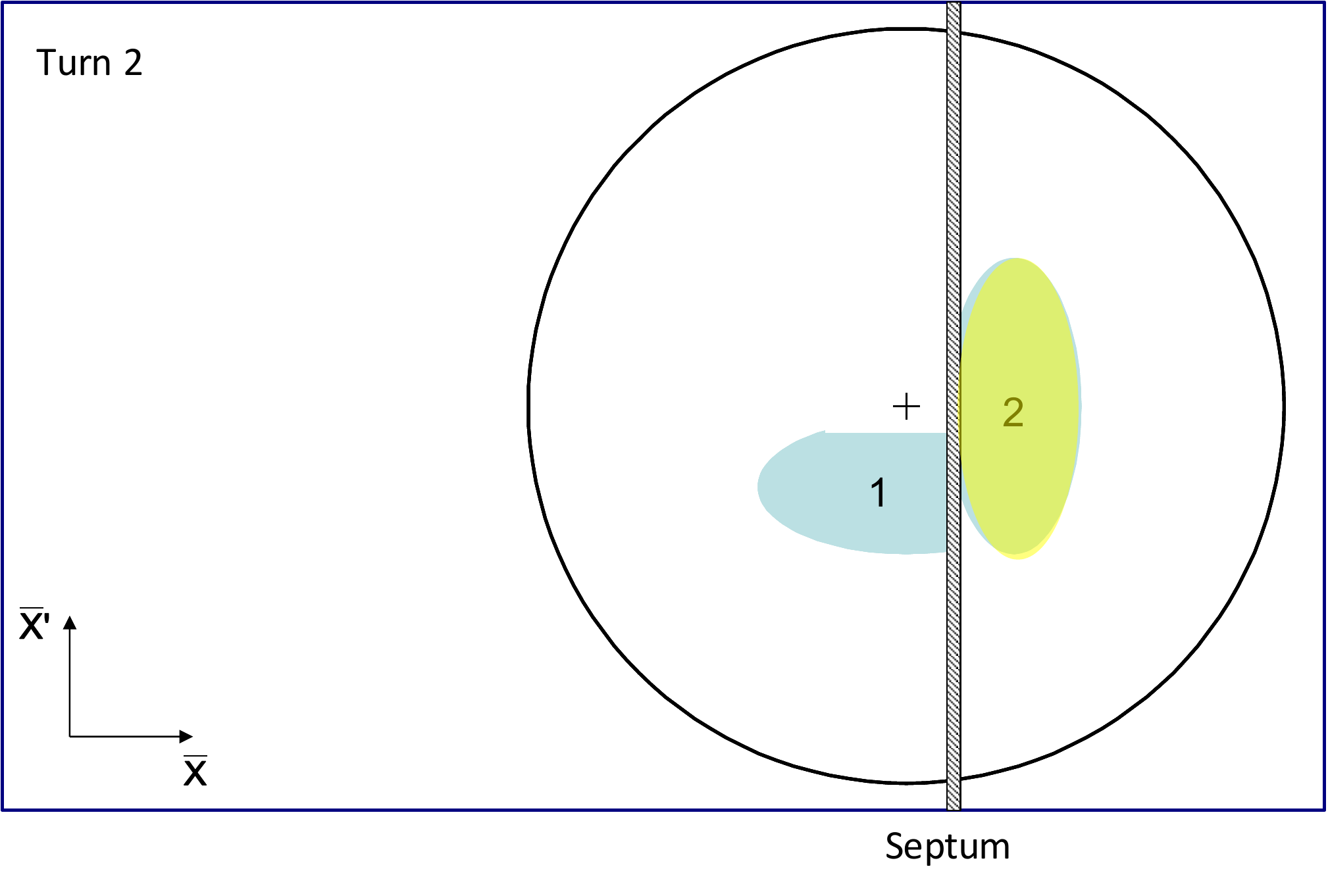}
\includegraphics[width=\myfigscale\linewidth]{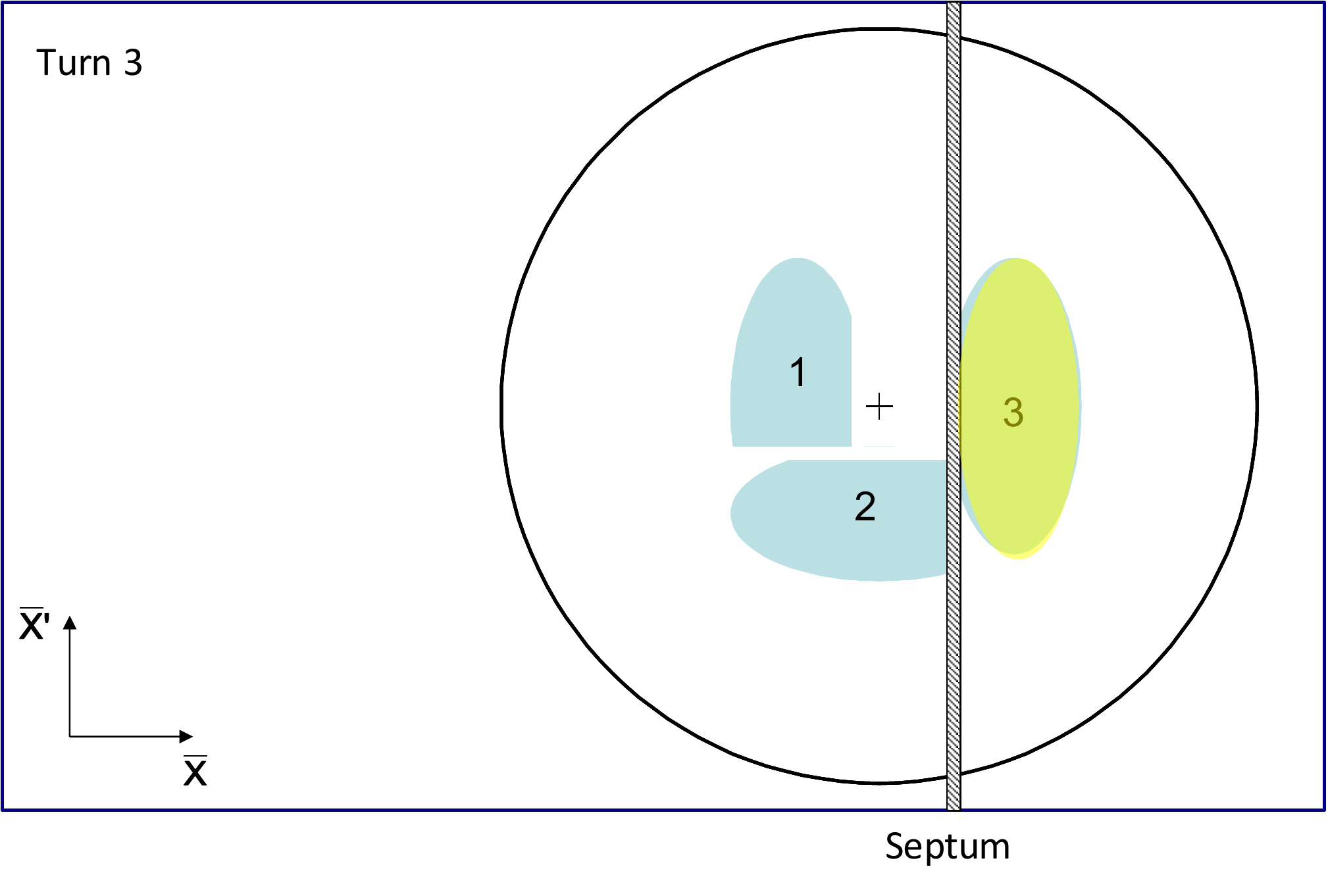}
\includegraphics[width=\myfigscale\linewidth]{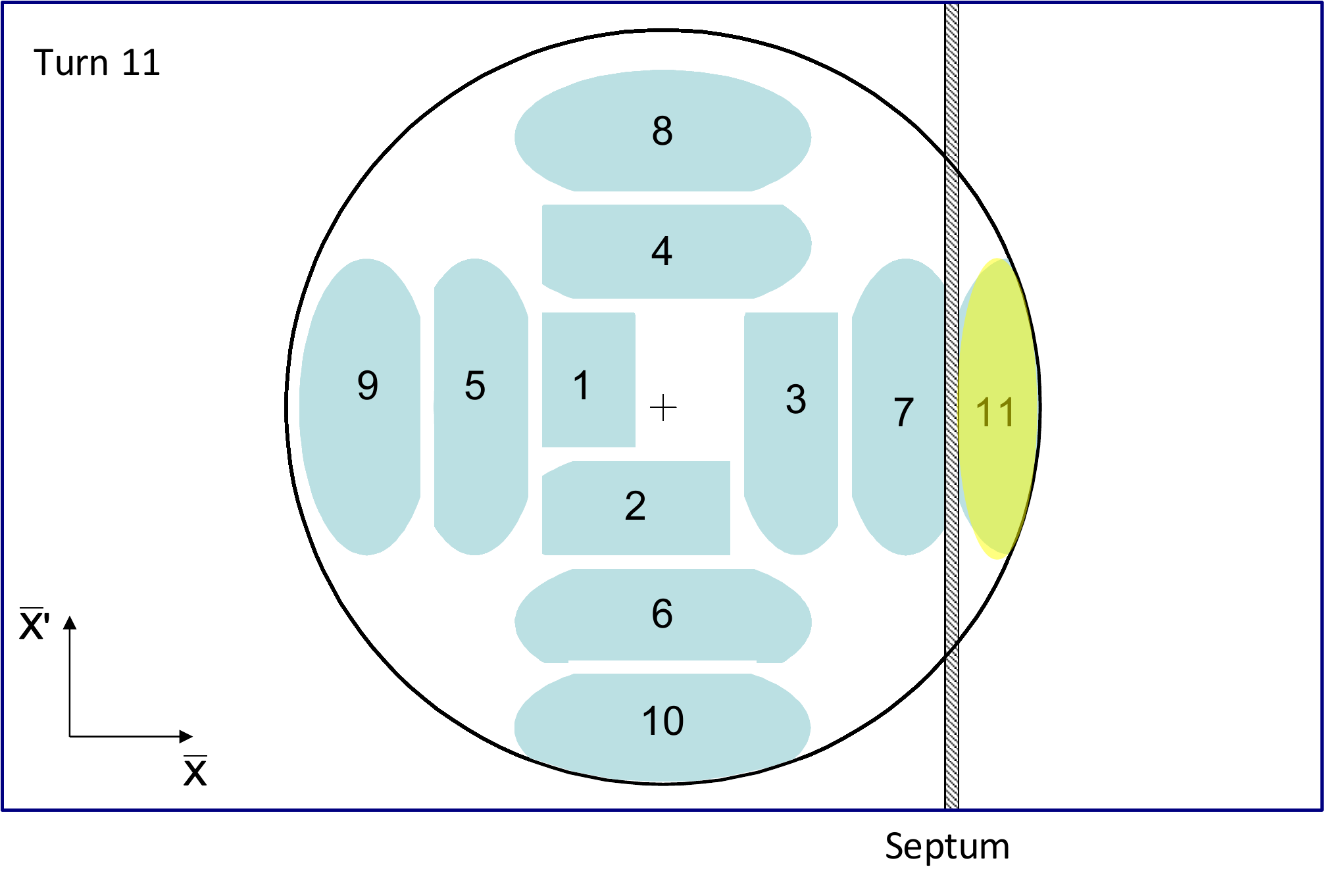}
\includegraphics[width=\myfigscale\linewidth]{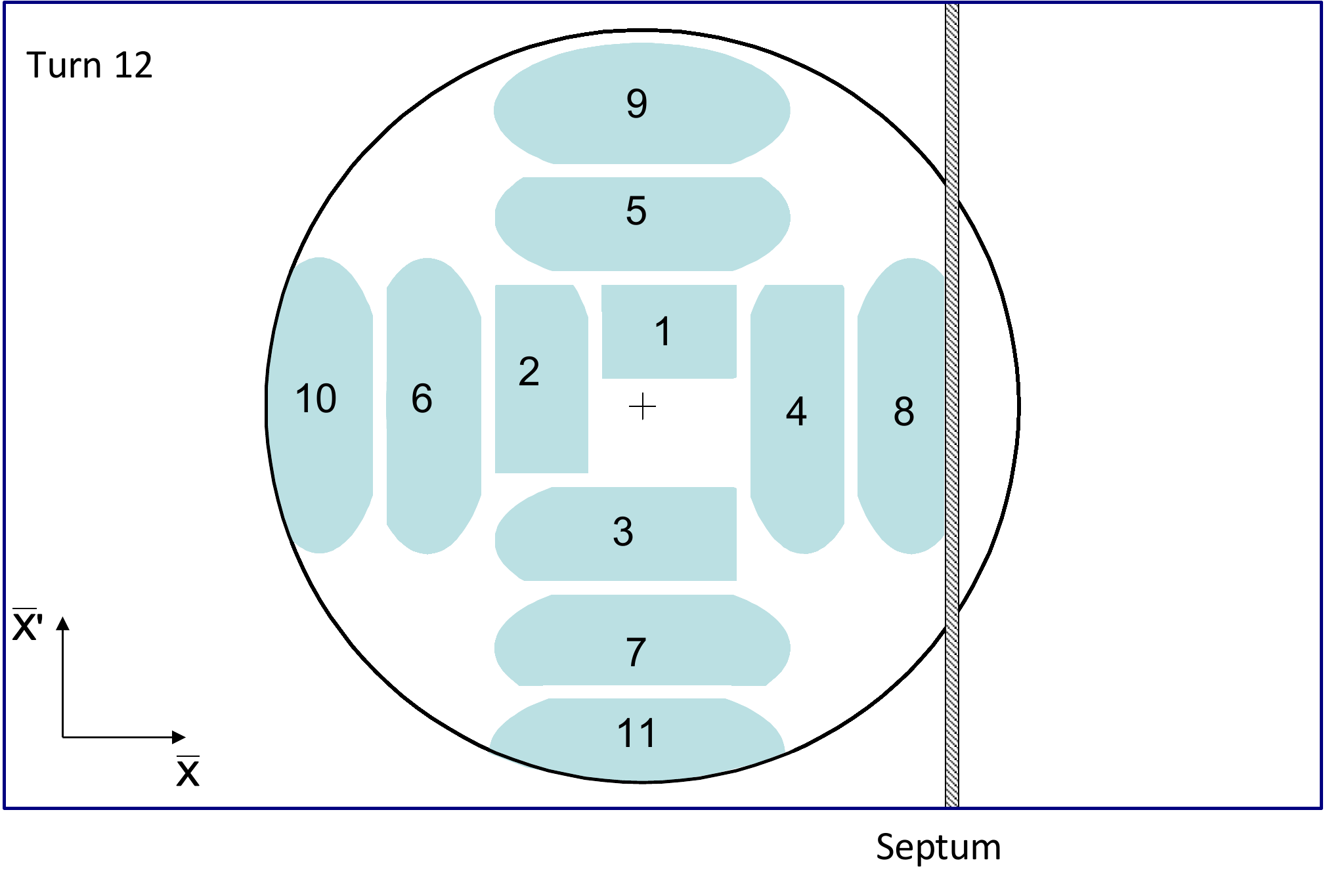}
\includegraphics[width=\myfigscale\linewidth]{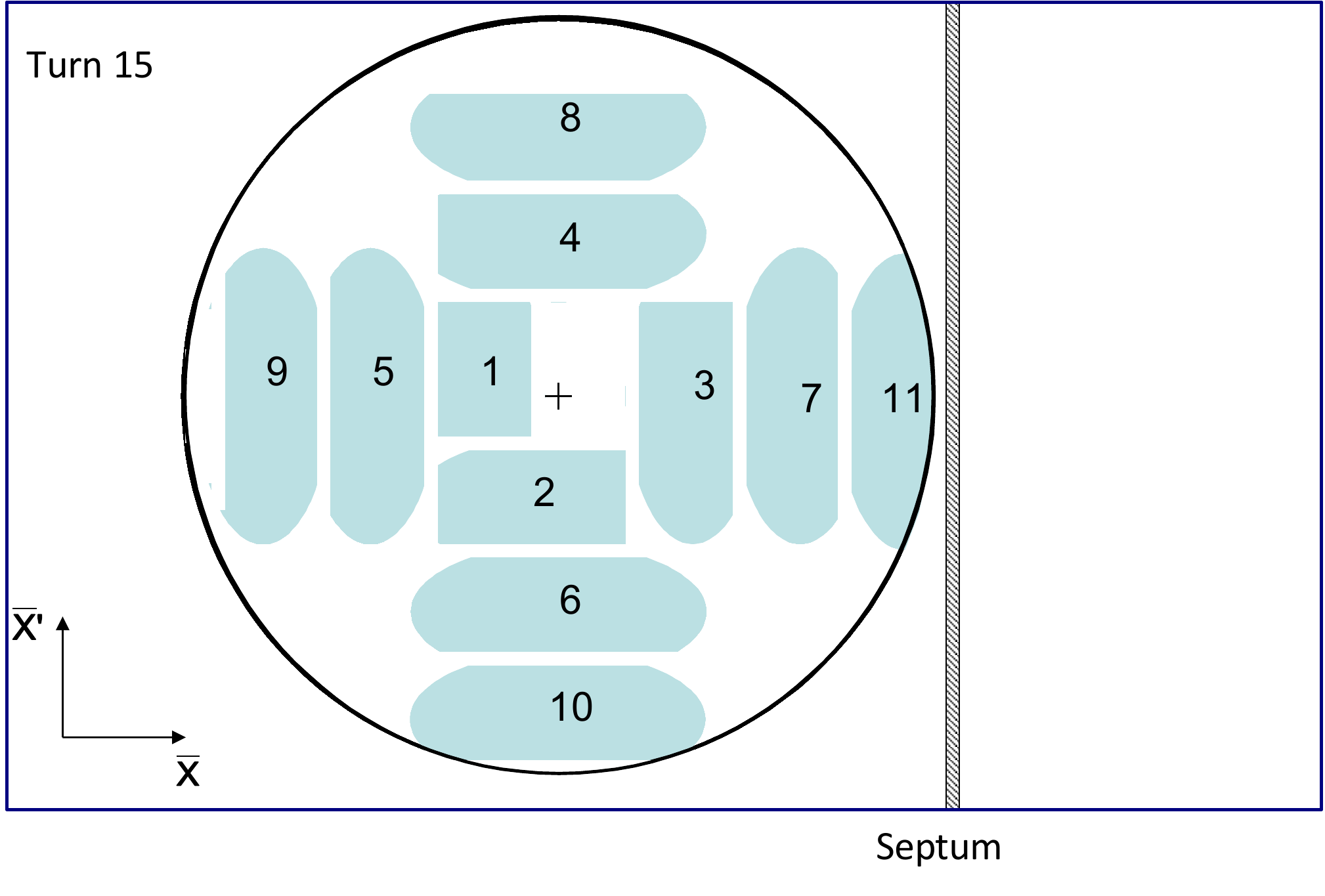}
\caption{Transverse painting in normalised phase space over fifteen turns assuming a fractional tune $Q_h$=0.25.
The injected pulse is shown in green. Note that the closed orbit (indicated by the +) moves to the left away from the septum over the different turns.}
\label{PhaseMulti}
\end{figure}

The multi-turn injection is performed by the combination of a septum with a set of three or four bumpers
(dipole magnets), as shown in Fig.\,\ref{fig:multiturn}. 
The bumpers initially bring the orbit close the the septum blade, and their kick is reduced in time in a programmed way,
such that the first beam occupies the central region and the later bunches the periphery
of the transverse phase space acceptance (Fig.\,\ref{PhaseMulti}). 

Filamentation, often space-charge driven, will at the end lead to a quasi-uniform charge density.
The finite thickness of the septum blade makes it necessary to keep a certain distance of the different 
beamlets of the injected beam, so that the resulting emittance $\varepsilon_r$ in the receiving machine is:
\begin{equation}
\varepsilon_r \gtrsim 1.5 \, N \, \varepsilon_i \label{eq:emittance}
\end{equation}
where $\varepsilon_i$ is the emittance of the injected beam and $N$ the number of injected pulses.
Even though multi-turn injection is essential to accumulate high intensity at low energy when space charge 
plays a dominant effect, it presents several disadvantages. 
The width of the septum blade reduces the available aperture, leading to high losses (up to 30\%\,-\,40\%)
from the circulating beam hitting the septum, and due to the limited acceptance,
the injection can be performed over a maximum of 10\,-\,20 turns.

\subsection{Charge exchange \texorpdfstring{H$^-$}{H-} injection}
An elegant alternative to avoid the mentioned disadvantages of conventional multi-turn injection is 
charge exchange injection~\cite{bib:chiara,bib:chiara2}. In particular, charge exchange injection overcomes the limitation
of not being able to inject into an already populated area of the transverse phase phase.

\begin{figure}[!b]
\centering\includegraphics[width=0.8\linewidth]{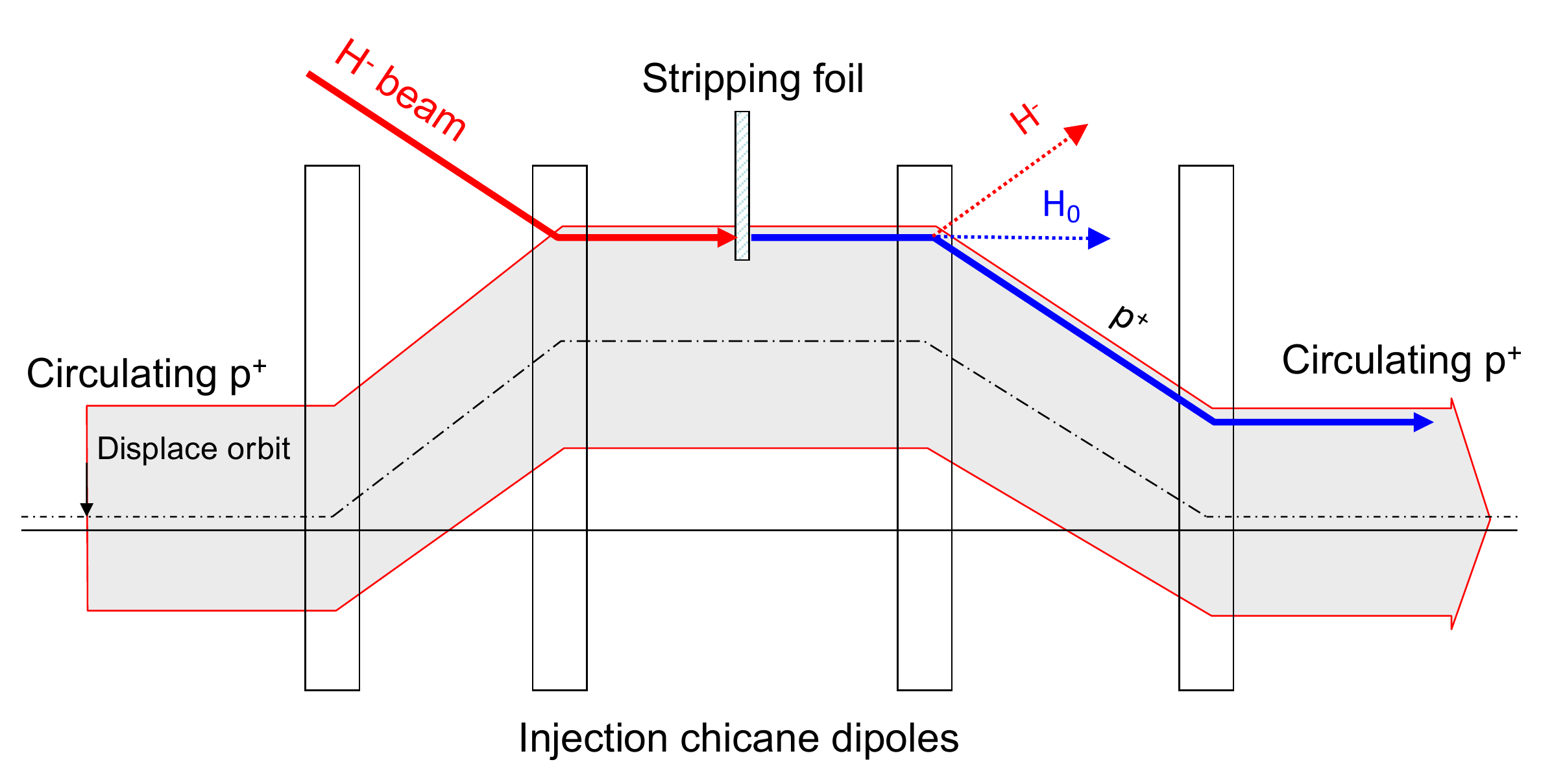}
\caption{Principle of H$^-$ charge exchange injection. A magnetic chicane, together with a thin stripping foil, allows to
inject the beam into an already populated area of the transverse phase space.}
\label{fig:charge-ex}
\end{figure}

It consists of converting H$^-$ ions into protons by using a thin stripping foil inside a magnetic chicane, 
allowing injection into the same phase space area, as the negatively charged H$^-$ ions can be brought
onto the same orbit at the foil as the already circulating protons (see Fig.~\ref{fig:charge-ex}).    
Another magnet chicane can be used to create a decaying closed orbit bump and 
paint uniformly the beam in the transverse phase space.
The thickness of the stripping foils has to be carefully chosen to maximise the stripping efficiency while minimising emittance
blow-up and losses. Typical foils are made of Carbon and have a thickness of 50\,-\,200\,$\upmu$g/cm$^{2}$ for energies
varying between 50\,MeV and 800\,MeV. The chicane is switched off at the end of the injection, to avoid excessive foil heating and
further emittance blow-up.
In some cases, longitudinal phase space painting can be performed by varying the energy of the injected 
beam turn-by-turn by scaling the voltage of the linac accordingly. 
A chopper system in the linac is used to match the length of the injected batch to the bucket.

The absence of the septum and the nature of this method allows reducing the injection losses to a few percent. 
Moreover the process can continue over tens up to hundred turns allowing to accumulate significant
charge densities and produce a higher brightness beam. Also in this case, the final
particle distribution is determined by the painting and the filamentation due to space charge. 

\subsection{Lepton injection}
The big difference of leptons with respect to hadrons is that the oscillations are strongly damped by
synchrotron radiation damping~\cite{bib:sr}. So injection errors are much less critical, as they will not lead to
an emittance growth like in the hadron case.

\begin{figure}[!b]
\centering\includegraphics[width=0.8\linewidth]{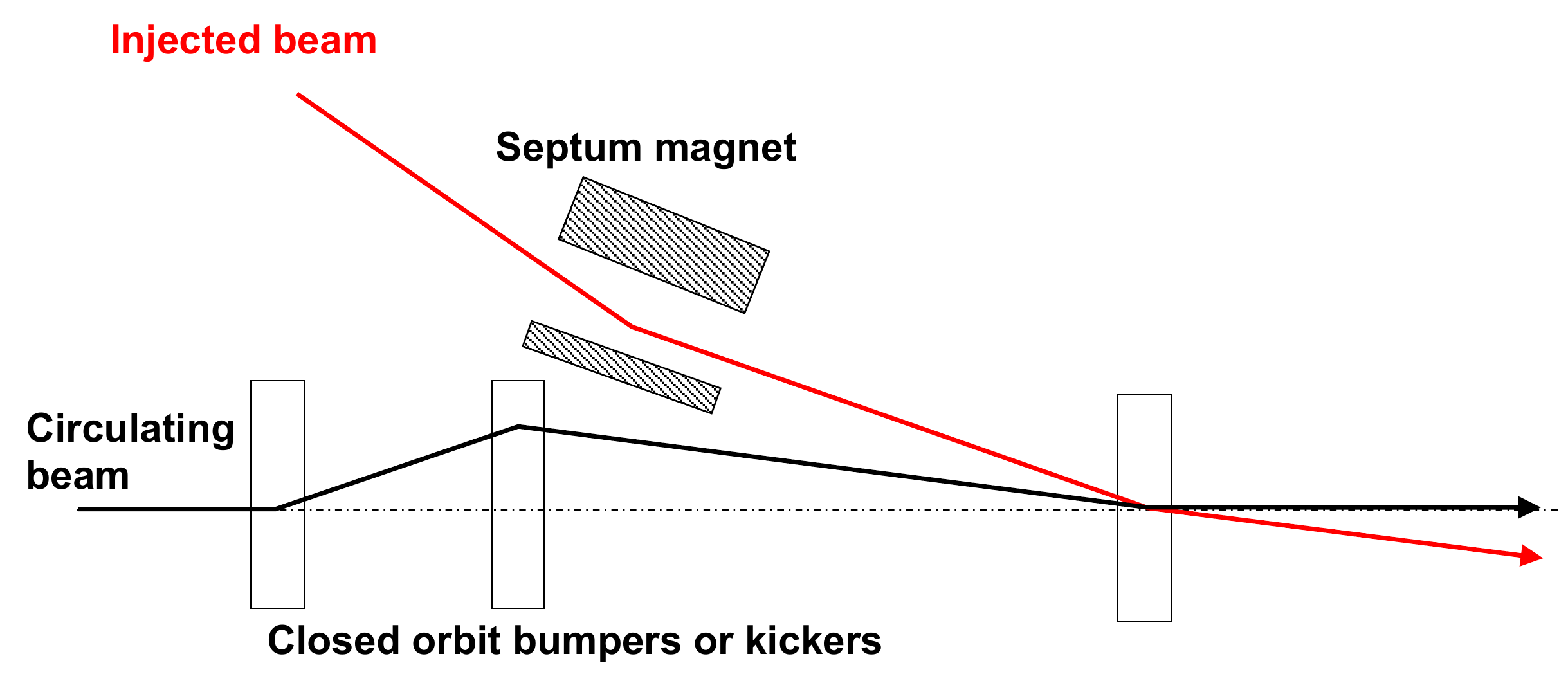}
\caption{Principle layout of betatron injection for leptons.}
\label{fig:betatron}
\end{figure}

The injection scheme can be similar to the hadron injection with kickers or bumpers (see Fig.~\ref{fig:betatron}).
The beam is injected with an angle with respect to the closed orbit. So the
injected beam performs damped betatron oscillations about the closed orbit and will finally merge with
the already circulating beam, asymptotically reaching the equilibrium emittance of the receiving accelerator.
Like this, a larger intensity can be accumulated in the same phase space area. As this method relies on the damping
of the betatron oscillations, it is also called 'betatron injection'.

\begin{figure}[!b]
\centering\includegraphics[width=0.8\linewidth]{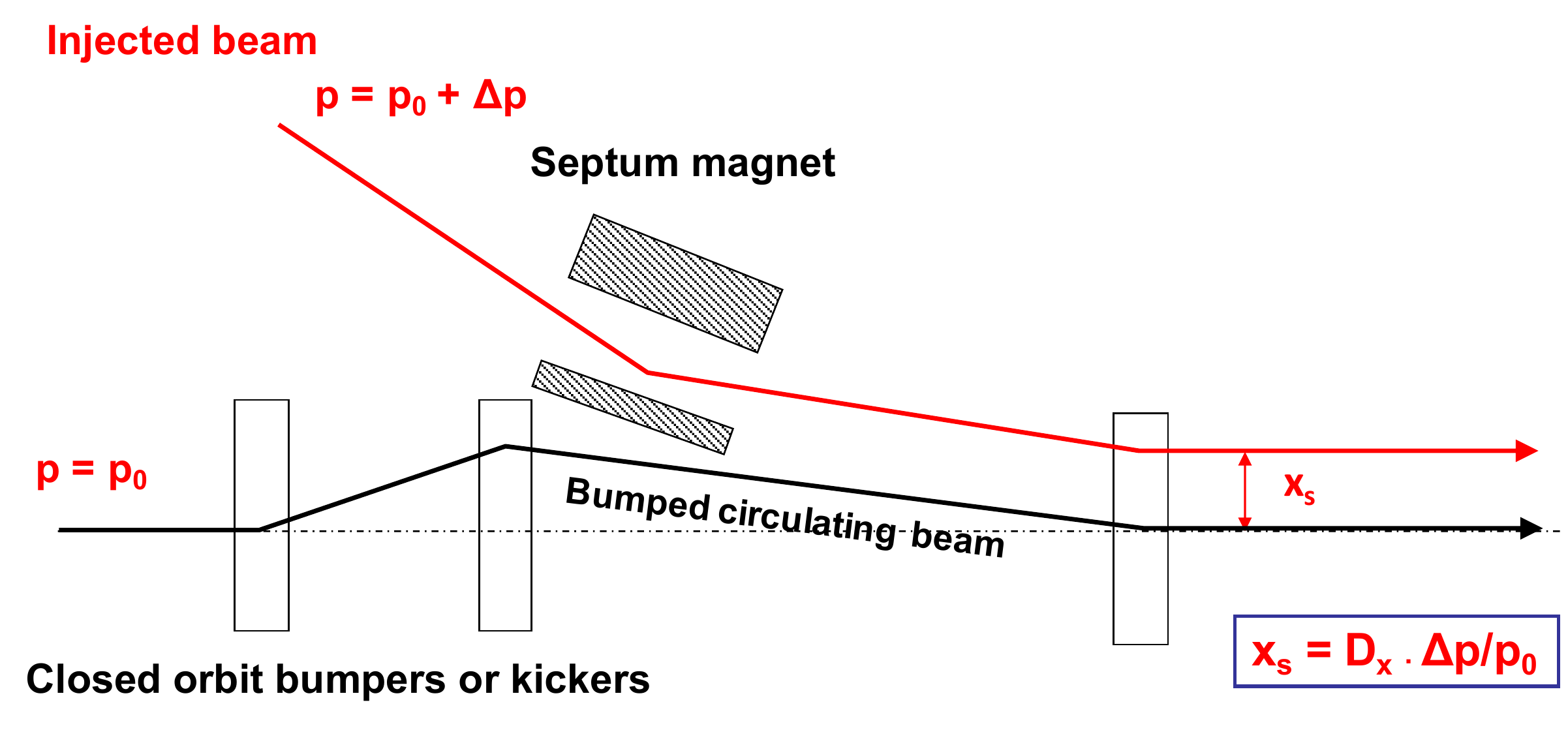}
\caption{Principle of synchrotron injection. The beam is injected with an energy offset at a location with dispersion.}
\label{fig:synchrotron}
\end{figure}

The damping in the longitudinal phase space is typically twice faster than in the transverse. This fact can be used
to derive the synchrotron injection scheme with this faster damping, shown in Fig.~\ref{fig:synchrotron}.
The beam is injected with a given momentum difference $\Delta p$ with respect to the beam circulating
with a momentum of $p_0$.
The beam is injected parallel to the circulating beam, onto a dispersion orbit for a particle having the momentum
$p = p_0 + \Delta p/p_0$. The injected beam needs to have the position $x_s = D_x\,\Delta p/p_0$, where
$D_x$ is the dispersion function at this location.
The injected beam makes damped synchrotron oscillations but does not perform betatron oscillations.
So the orbit difference with respect to the closed orbit will be $x(s) = D_x(s) \Delta p/p_0$, which vanishes
for locations with zero dispersion. As the beam optics around physics experiments is often designed to have
vanishing dispersion, the synchrotron injection will not affect the orbit there and will result in lower background
in the detectors from the injection process.

\section{Extraction principles}
Depending on the requirements, different extraction techniques exist:
\begin{itemize}
\item Fast extraction, $\leq 1$ turn
\item Non-resonant (fast) multi-turn extraction, few turns
\item Resonant low-loss (fast) multi-turn extraction, few turns
\item Resonant (slow) multi-turn extraction, many thousands of turns.
\end{itemize}
The extraction energy of the beam is usually higher than at injection and consequently requires
stronger elements to reach the necessary deflection.
So at high energies many kicker and septum modules may be needed.
In order to reduce kicker and septum strength, the beam can be moved near to the septum by a closed orbit bump.
With a higher energy, the beam size is smaller than at injection, as it scales with $1/\sqrt{\gamma_r}$ (with $\gamma_r$ the relativistic gamma).

\subsection{Single-turn extraction}
\begin{figure}[!b]
\centering\includegraphics[width=0.8\linewidth]{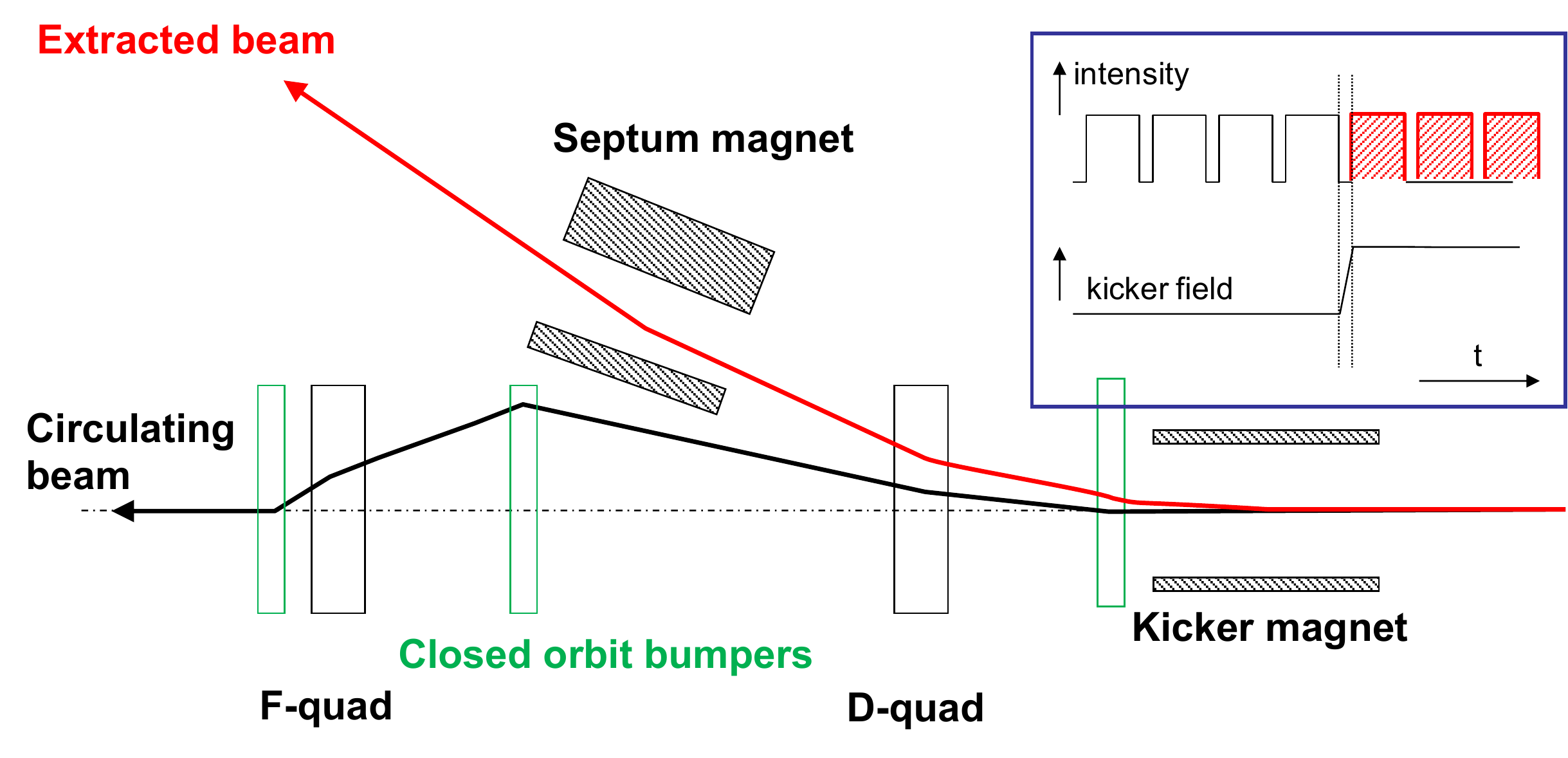}
\caption{Principle of single-turn extraction. }
\label{fig:extraction-single-turn}
\end{figure}

The single-turn extraction is basically a mirror image of the single-turn injection. The scheme is depicted in 
Fig.~\ref{fig:extraction-single-turn}.
A set of bumpers move the circulating beam close to septum to reduce kicker strength.
The kicker deflects the entire beam into the septum in a single turn. 
It is most efficient (with the lowest deflection angle required) for $90^\degree$ phase advance between kicker and septum.
The rise time of the kicker should be
short enough that it can reach its nominal deflection during a time when no beam is passing, otherwise part
of the beam will be lost. This might require to leave a gap in the longitudinal distribution of the bunches. 

Single-turn extraction is used for the transfer of beams between accelerators in an injector chain, for
secondary particle production (like neutrinos or radioactive beams) at a target, or for extracting the beam 
to a beam dump at the end of a fill in a storage ring (like the LHC).

\subsection{Non-resonant fast multi-turn extraction}
\begin{figure}[!b]
\centering\includegraphics[width=0.8\linewidth]{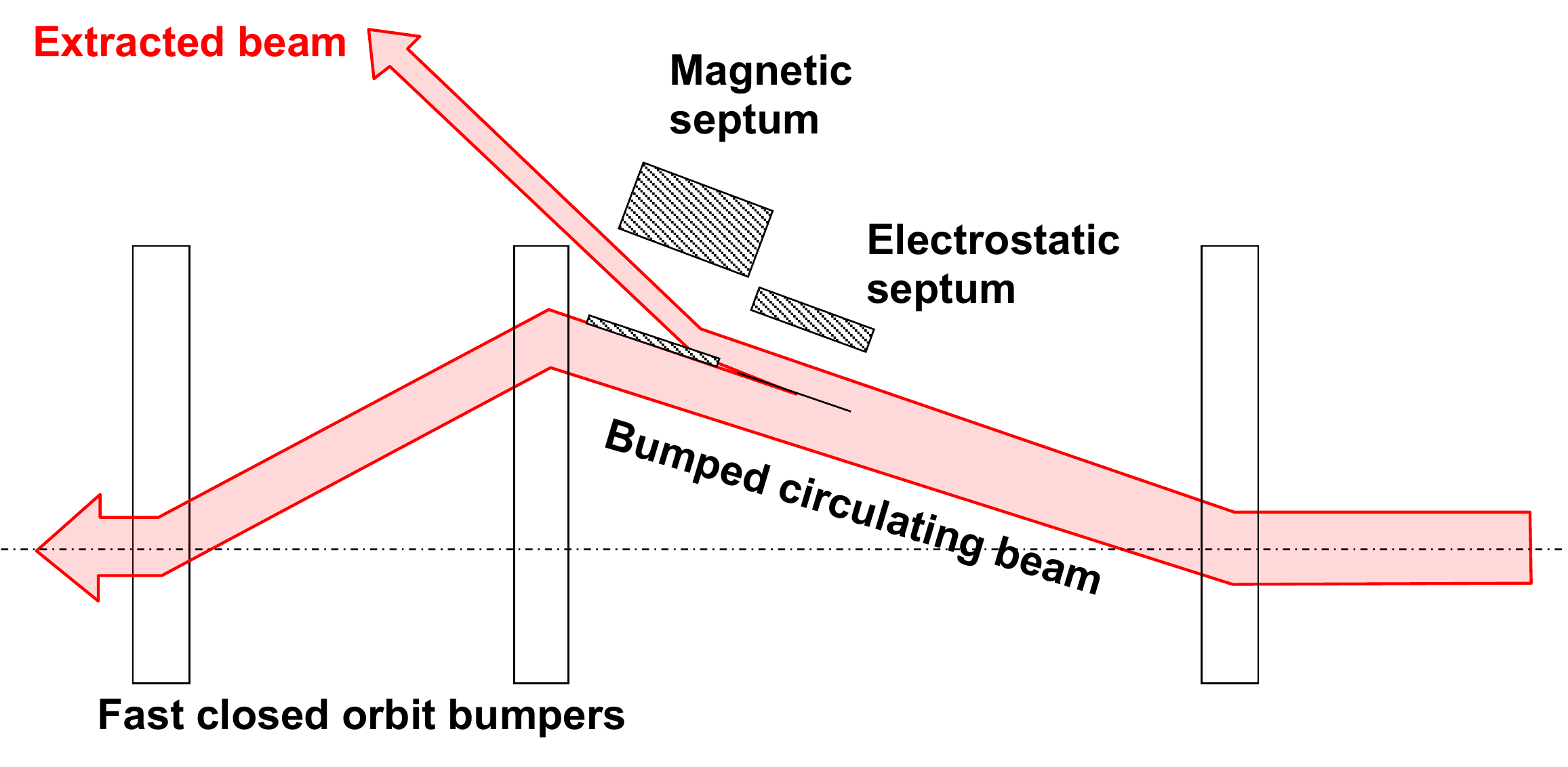}
\caption{Principle layout of non-resonant multi-turn extraction. }
\label{fig:extraction-multi-non-res}
\end{figure}

\begin{figure}[!bt]
1) \includegraphics[width=0.45\linewidth]{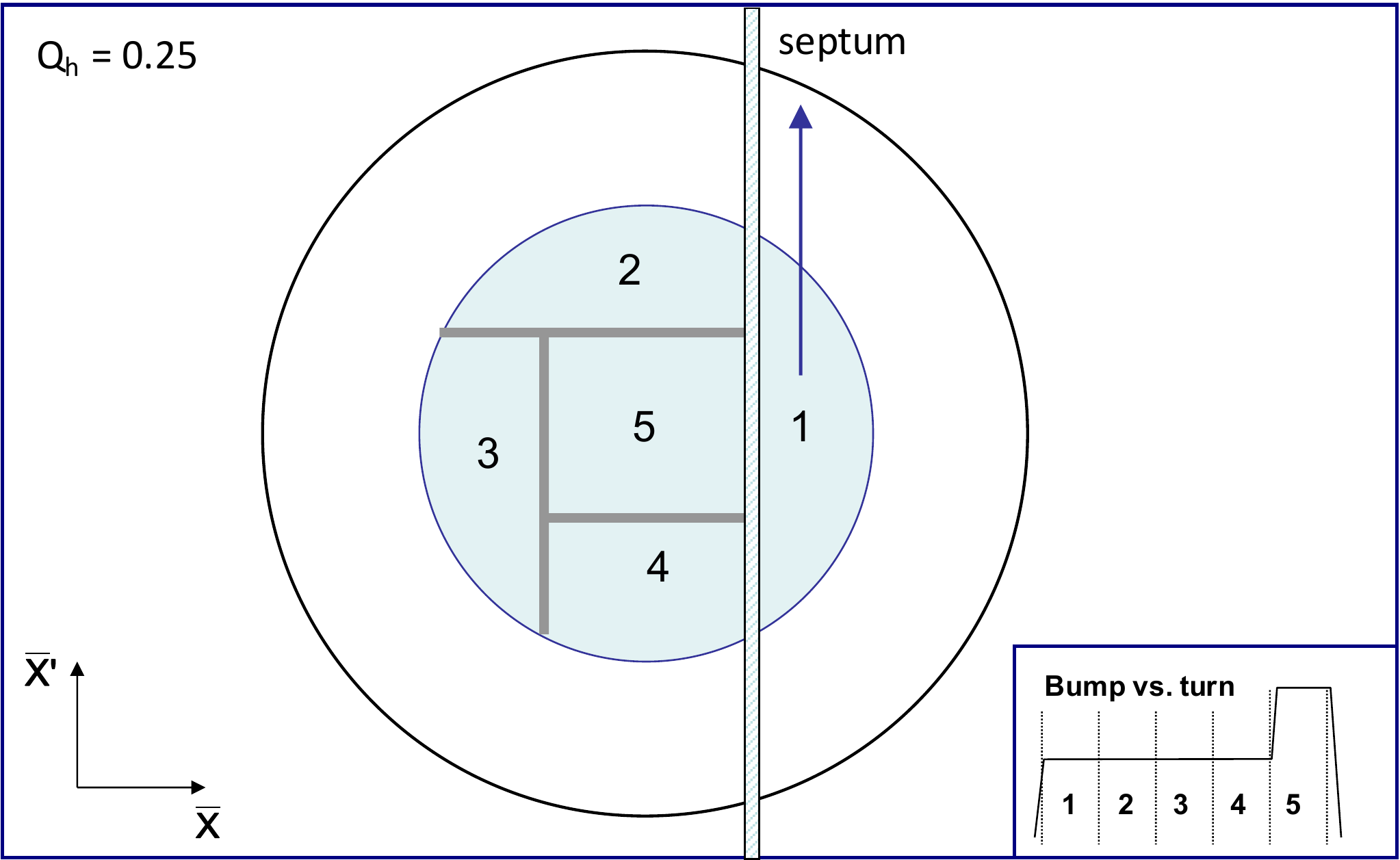}
2) \includegraphics[width=0.45\linewidth]{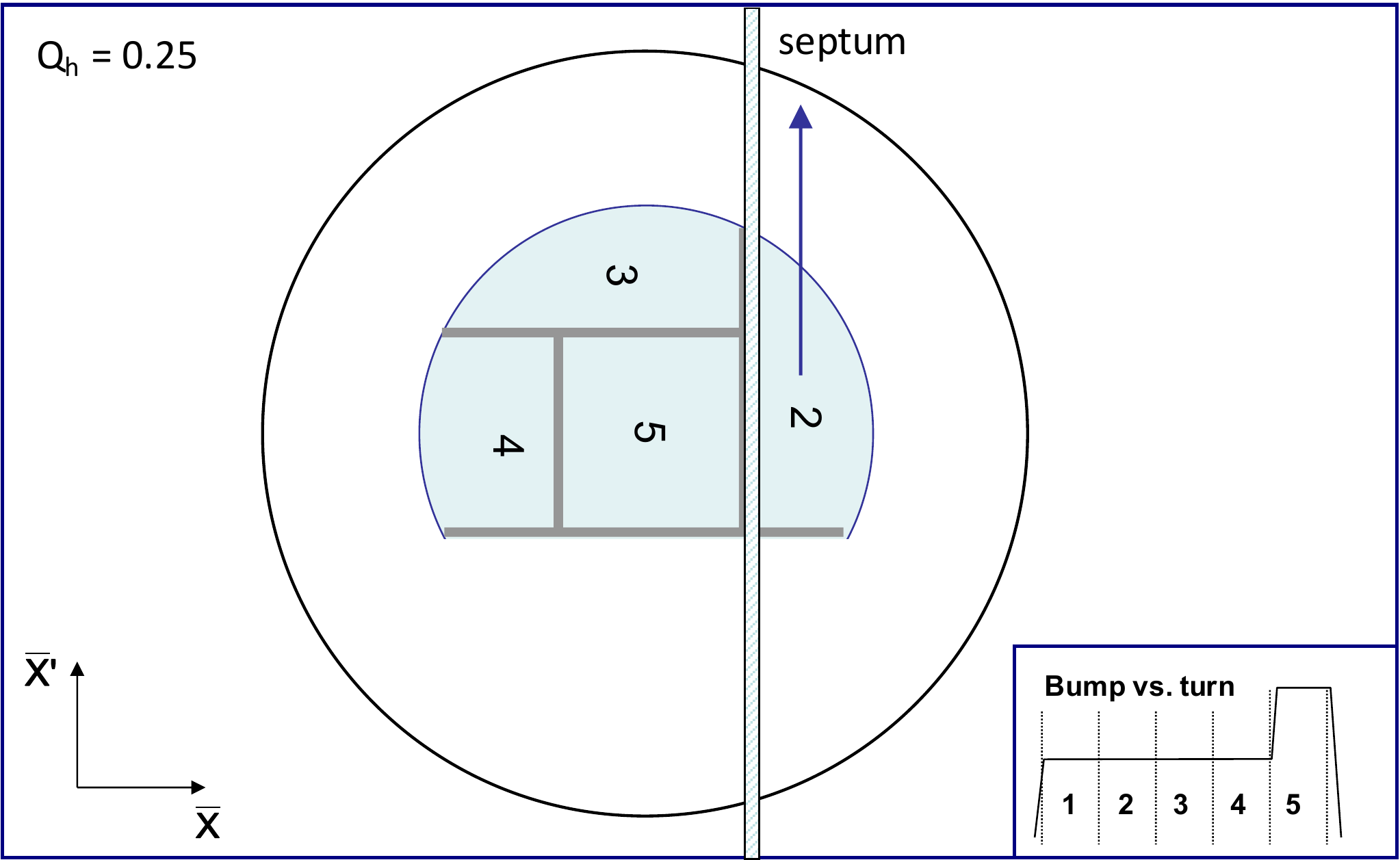}\\
3) \includegraphics[width=0.45\linewidth]{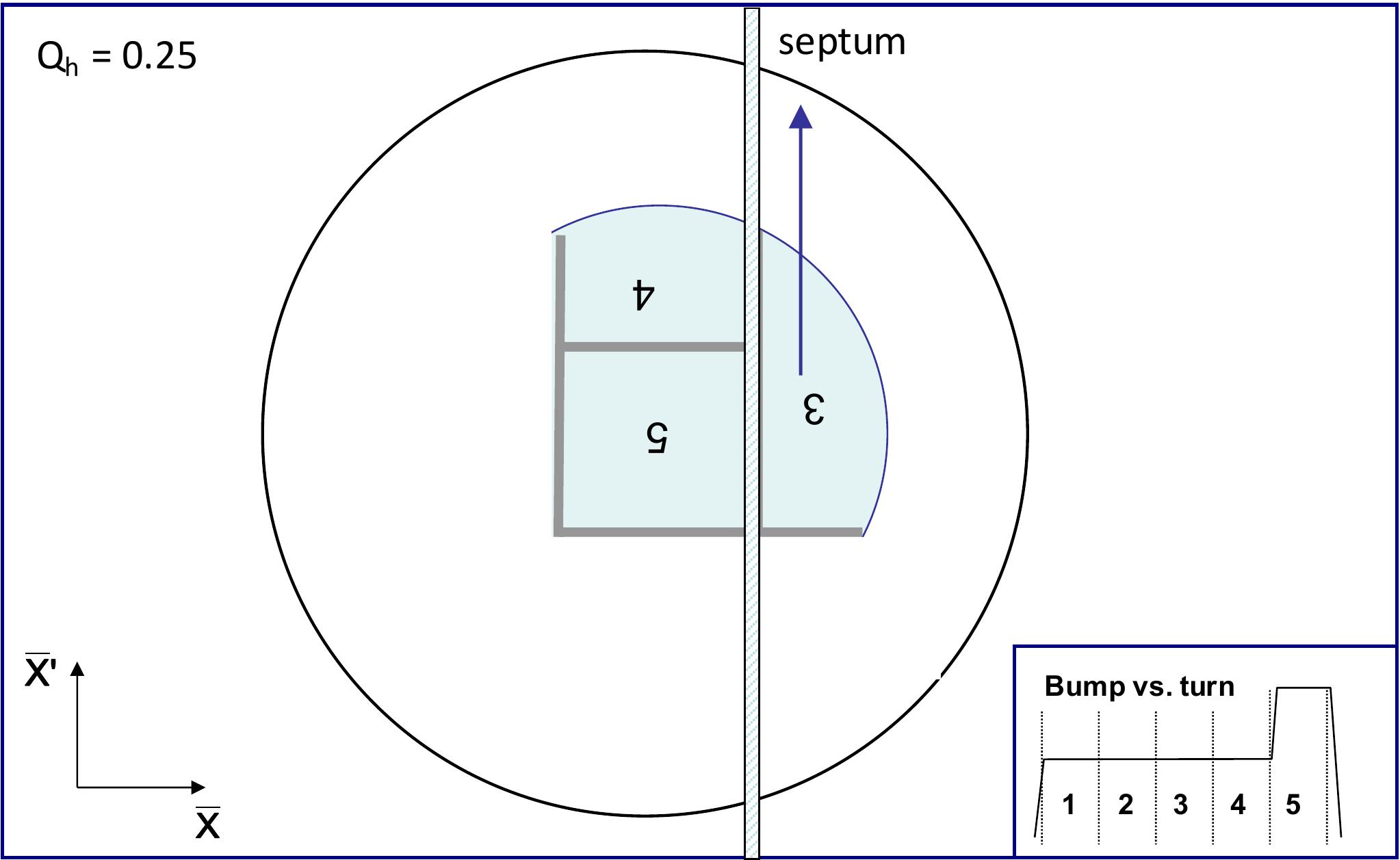}
4) \includegraphics[width=0.45\linewidth]{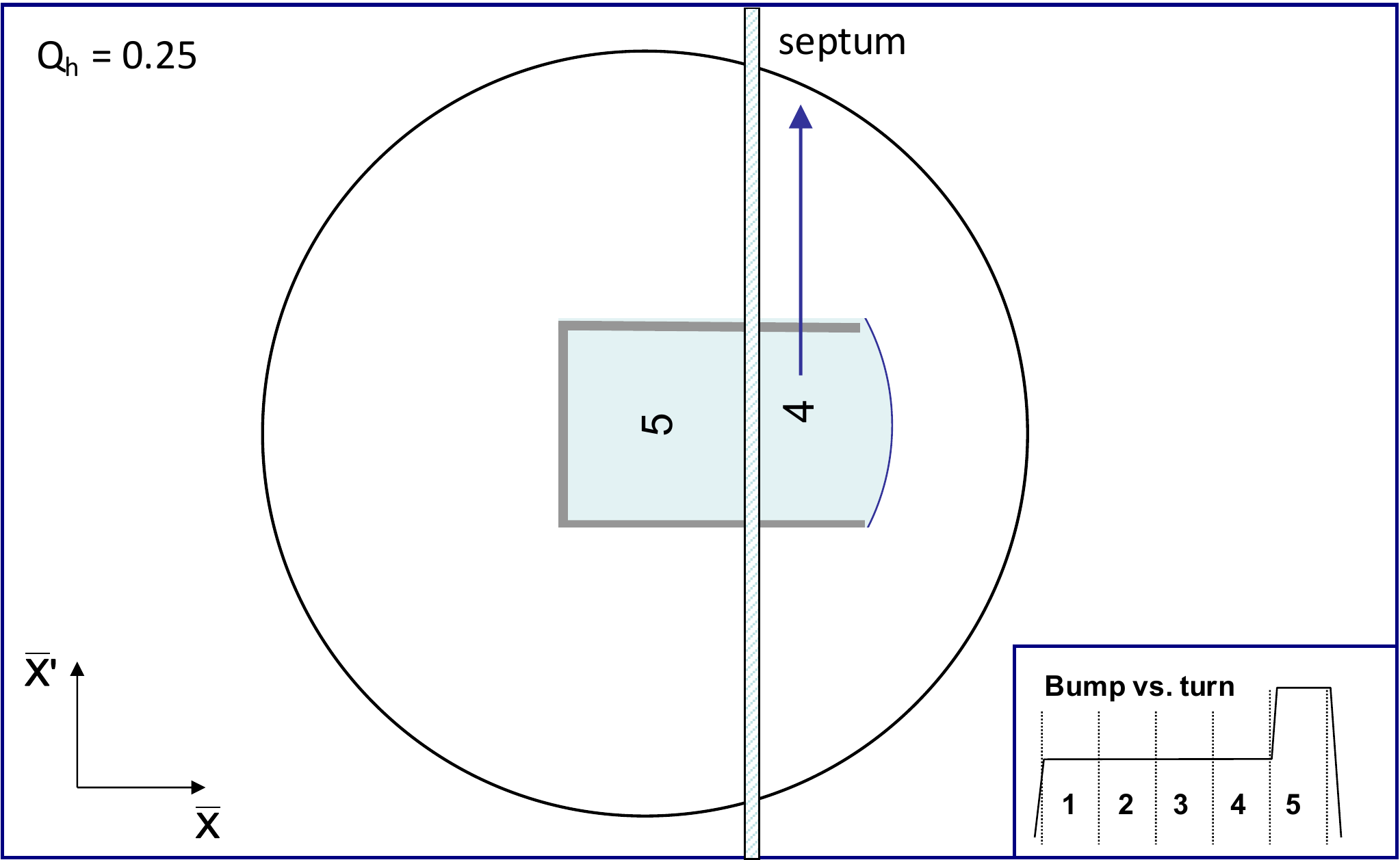}\\
5) \includegraphics[width=0.45\linewidth]{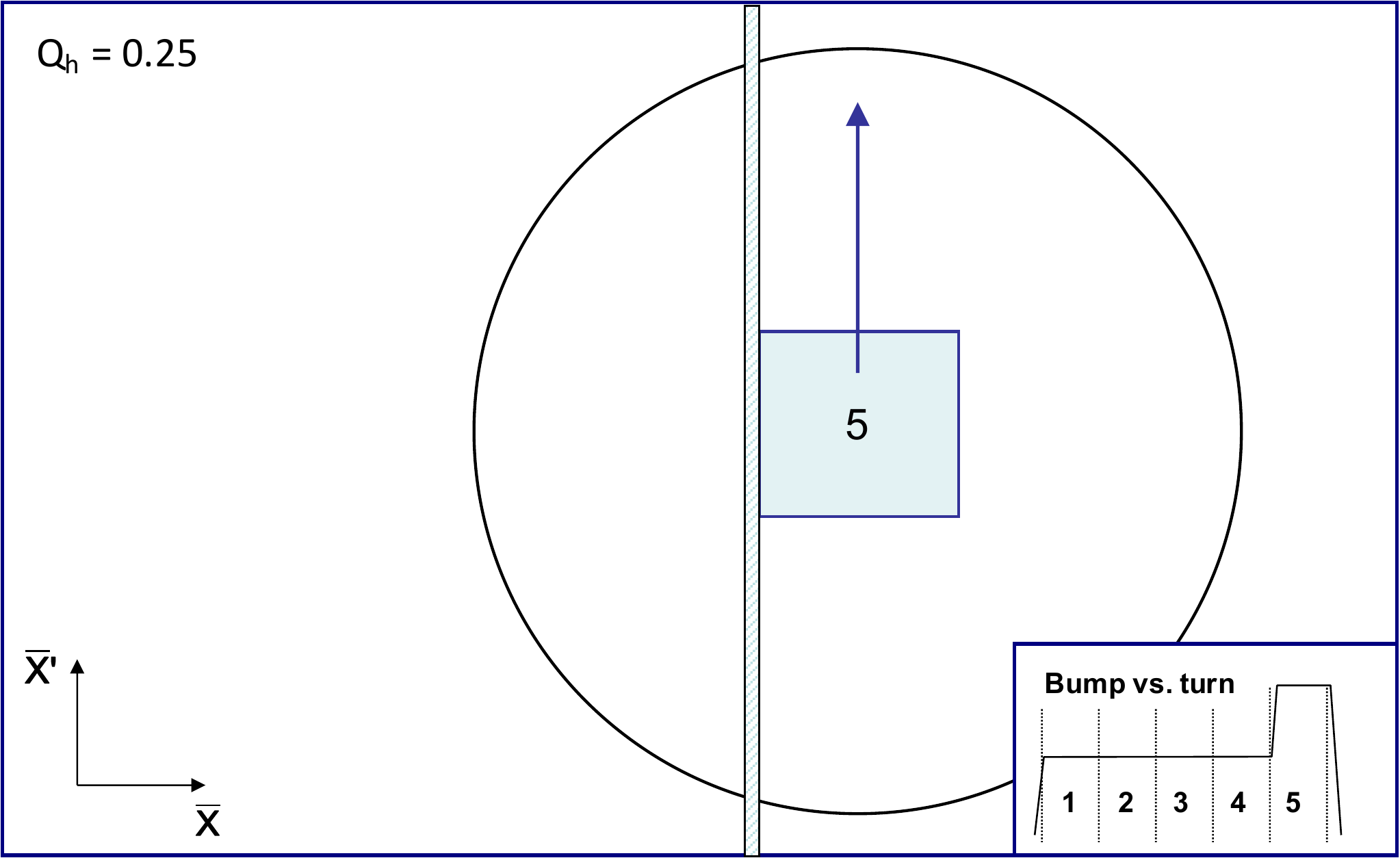}\hfill
\caption{Normalised phase space evolution of non-resonant multi-turn extraction for a tune of $Q_h = 0.25$
for five successive turns.
Note that the bump amplitude needs to be increased to extract the core of the beam in the fifth turn. }
\label{fig:extraction-ct}
\end{figure}

Some filling schemes require a beam to be injected in several turns into a larger machine. 
In this case, a part of the beam can be extracted each turn, leading to a longer beam pulse after the transfer.
Fig.~\ref{fig:extraction-multi-non-res} shows the principle layout and Fig.~\ref{fig:extraction-ct} shows the 
extraction process in normalised phase space. 
Fast closed orbit bumpers deflect the beam onto the septum, such that a part of the beam gets into
the extraction path.
The beam extracted in a few turns, with the machine tune rotating the beam in phase space.
This is intrinsically a high-loss process; a very thin septum is essential to minimise the losses.
Often very thin electrostatic septa are combined with magnetic septa.
Nevertheless, about 15\% of the beam was lost in the so-called Continuous Transfer (CT) from
the CERN Proton Synchrotron (PS) to the Super Proton Synchrotron (SPS) for SPS fixed target beam.
Furthermore, it is difficult to get equal intensities per turn, and one gets 
different trajectories and different emittances for each turn.

\subsection{Resonant fast multi-turn extraction}
\begin{figure}[!bt]
\centering\includegraphics[width=0.82\linewidth]{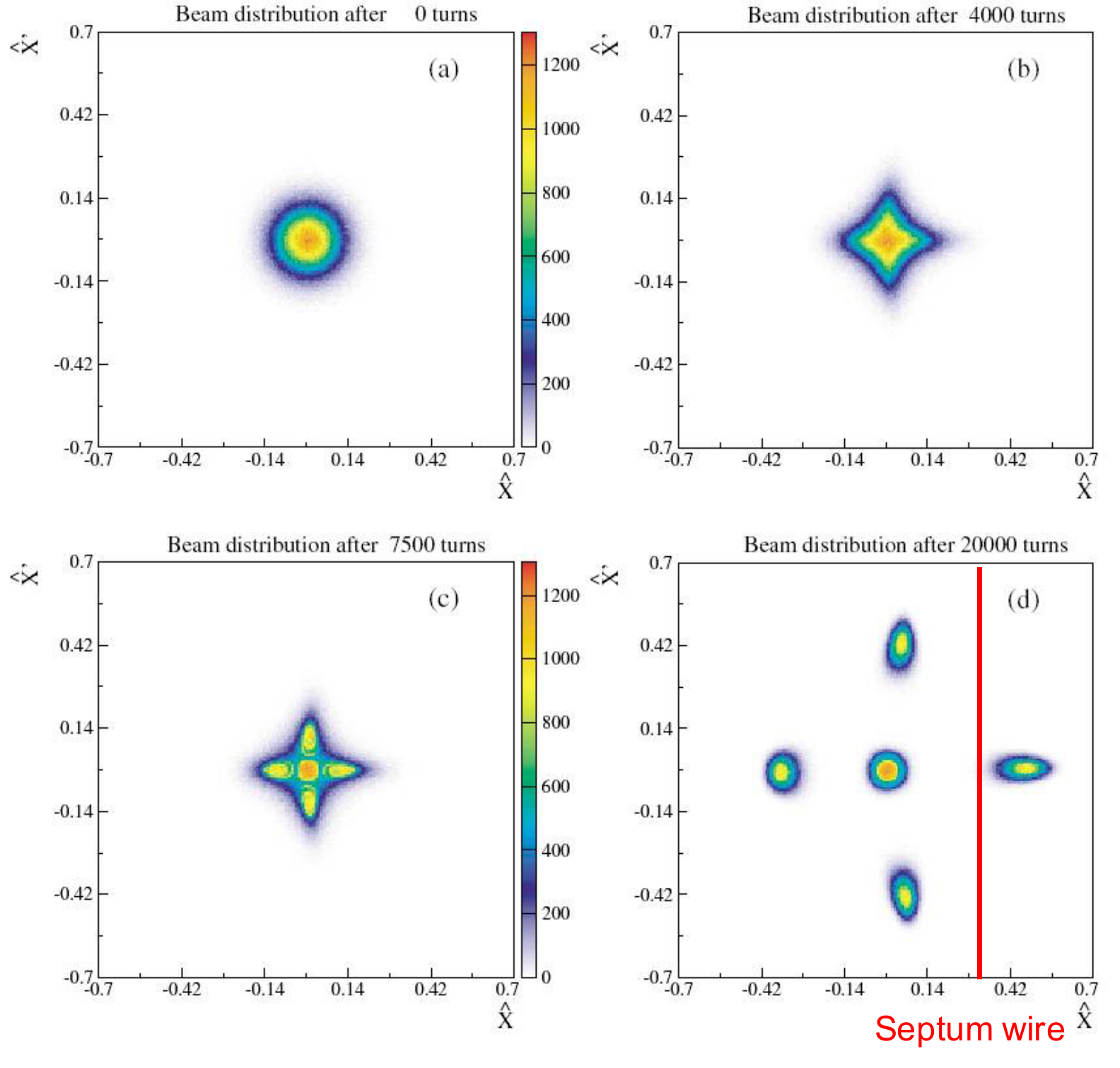}
\caption{Population of the resonant islands~\cite{bib:MTE-islands}. a) Unperturbed beam,
b) increasing non-linear fields,
c) beam captured in stable islands,
d) islands separated and beam kicked across septum. The beam is then extracted in five turns (also here the
core needs to be kicked by a stronger kick, similar to the non-resonant case).
}
\label{fig:MTE-islands}
\end{figure}

To overcome the aforementioned difficulties of the non-resonant multi-turn extraction, a resonant
fast multi-turn extraction was developed.
This method is based on adiabatic capture of part of the beam in stable “islands”, as illustrated in 
Fig.~\ref{fig:MTE-islands}. 
Non-linear fields (sextupoles and octupoles) create islands of stability in phase space.
A slow (adiabatic) tune variation is applied to cross a resonance and to drive particles into the islands (capture),
with the additional help of a transverse excitation (using the transverse damper in an excitation mode).
A variation of the multipole field strengths separates the islands in phase space further from the core,
such that the separation avoids losses on the septum blade (except during the rise time of the kicker
which deflects the beam into the extraction region of the septum).

So the losses are significantly reduced, and the phase space matching is improved with respect to the 
non-resonant multi-turn extraction, as the ‘beamlets’ have similar emittance and optical parameters.
This method replaced the earlier Continuous Transfer from the PS to the SPS.

\subsection{Resonant multi-turn slow extraction}
Also for this 'slow' extraction method, non-linear fields excite resonances, 
but here they are used to slowly (at the timescale
of up to several seconds) drive the beam across the septum. The principle layout of the extraction
is basically the same as in Fig.~\ref{fig:extraction-multi-non-res}, except that the orbit bumpers (which do not
need to be fast) just move the beam near the septum, the particles will be moved across the septum by
excitation of oscillations by the resonance.

The tune has to be adjusted close to an n$^\mathrm{th}$-order betatron resonance.
Corresponding multipole magnets are excited to define a stable area in phase space. 
When the stable area is slowly made smaller, the unstable particles get driven across the 
septum.

\begin{figure}[!bt]
a)\includegraphics[width=0.3\linewidth]{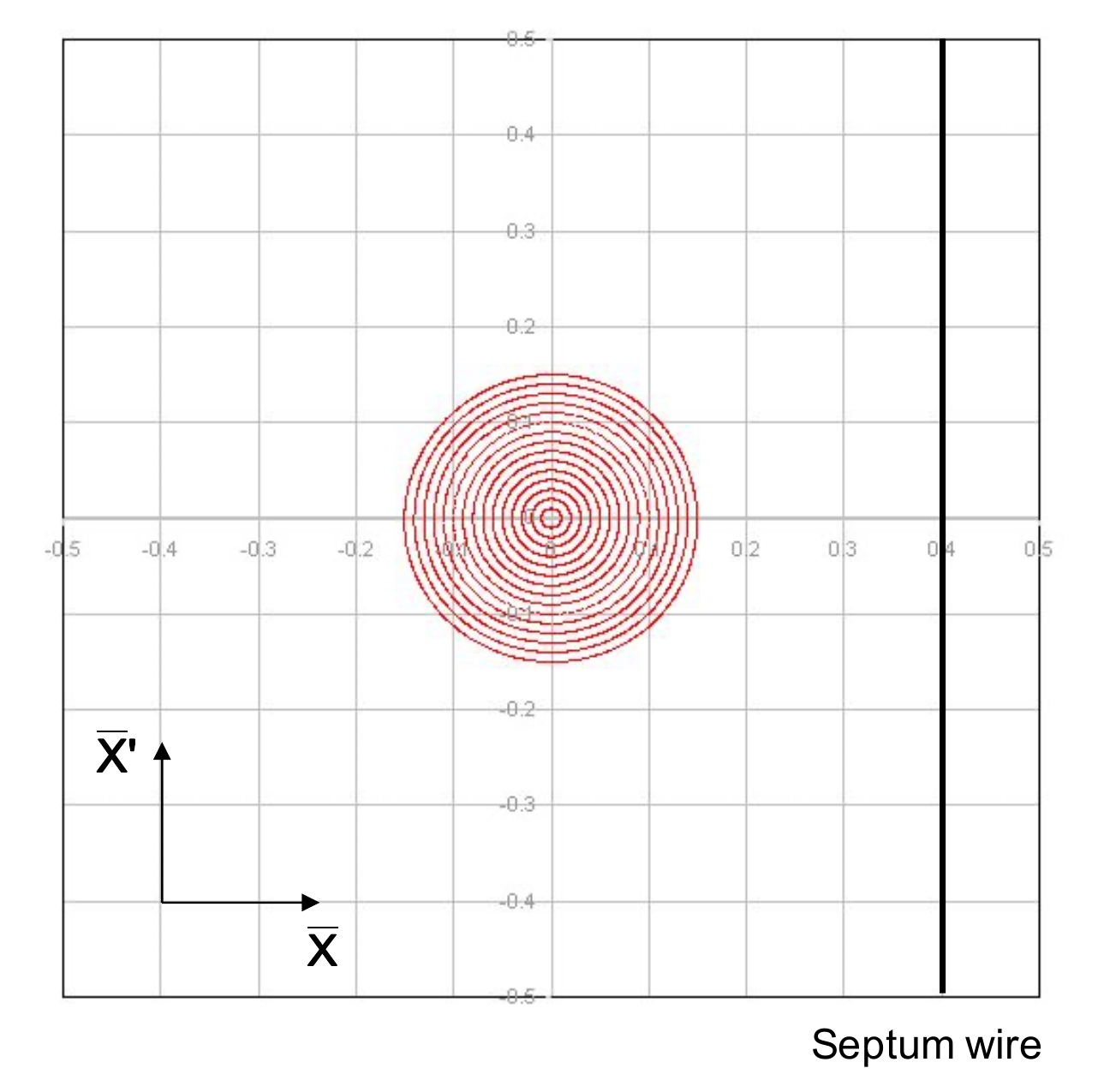}
b)\includegraphics[width=0.3\linewidth]{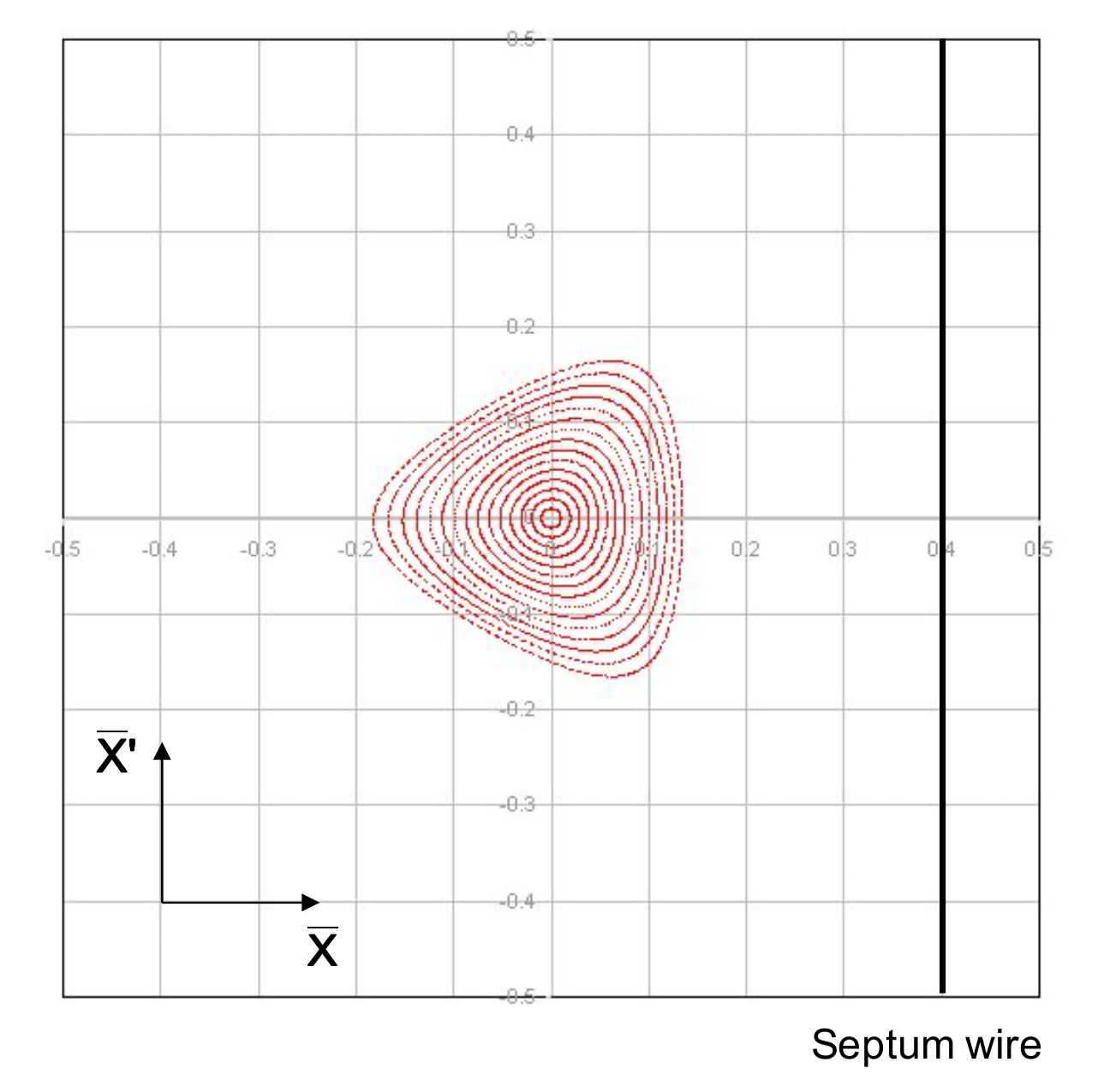}
c)\includegraphics[width=0.3\linewidth]{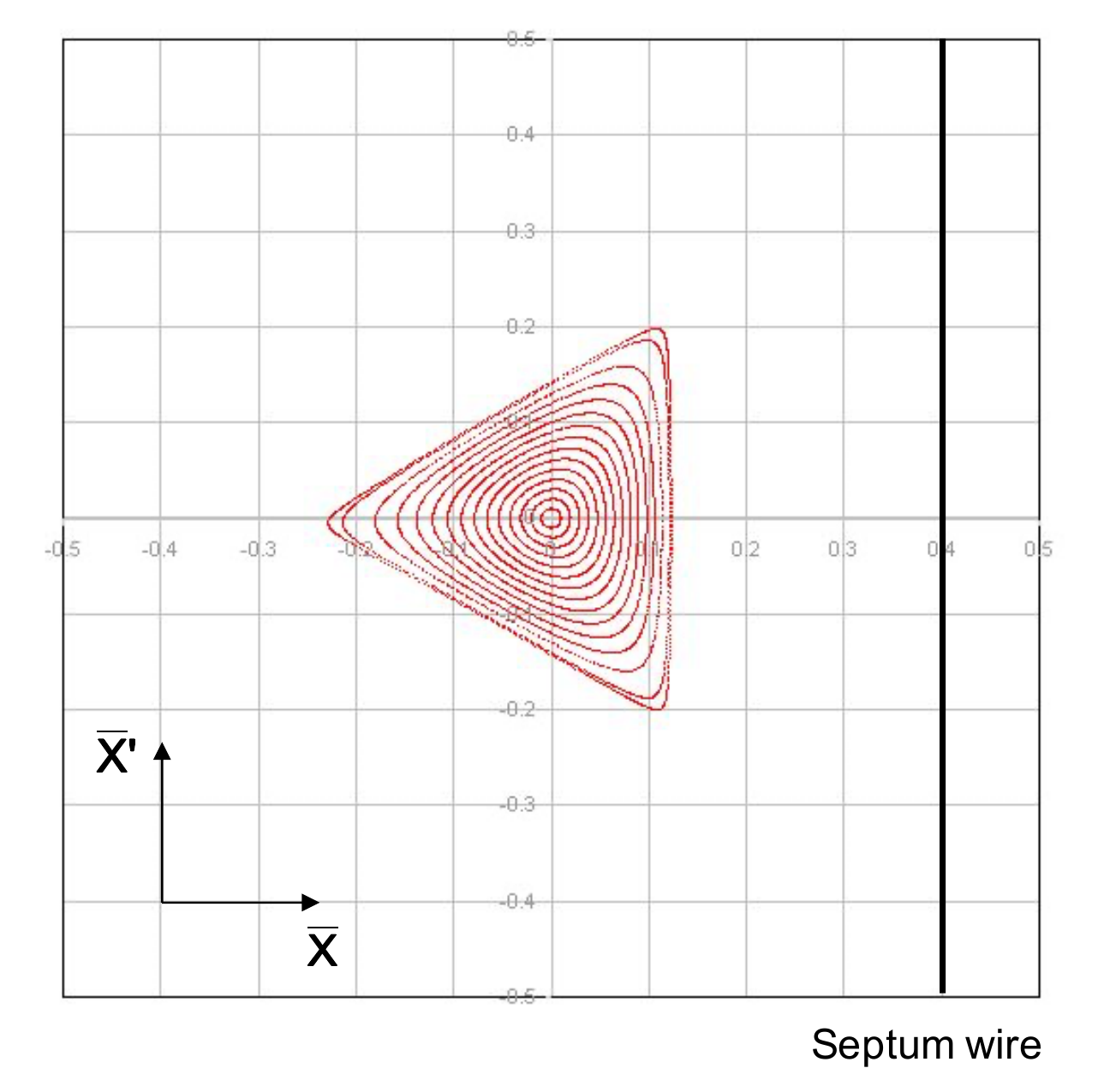}\\
d)\includegraphics[width=0.3\linewidth]{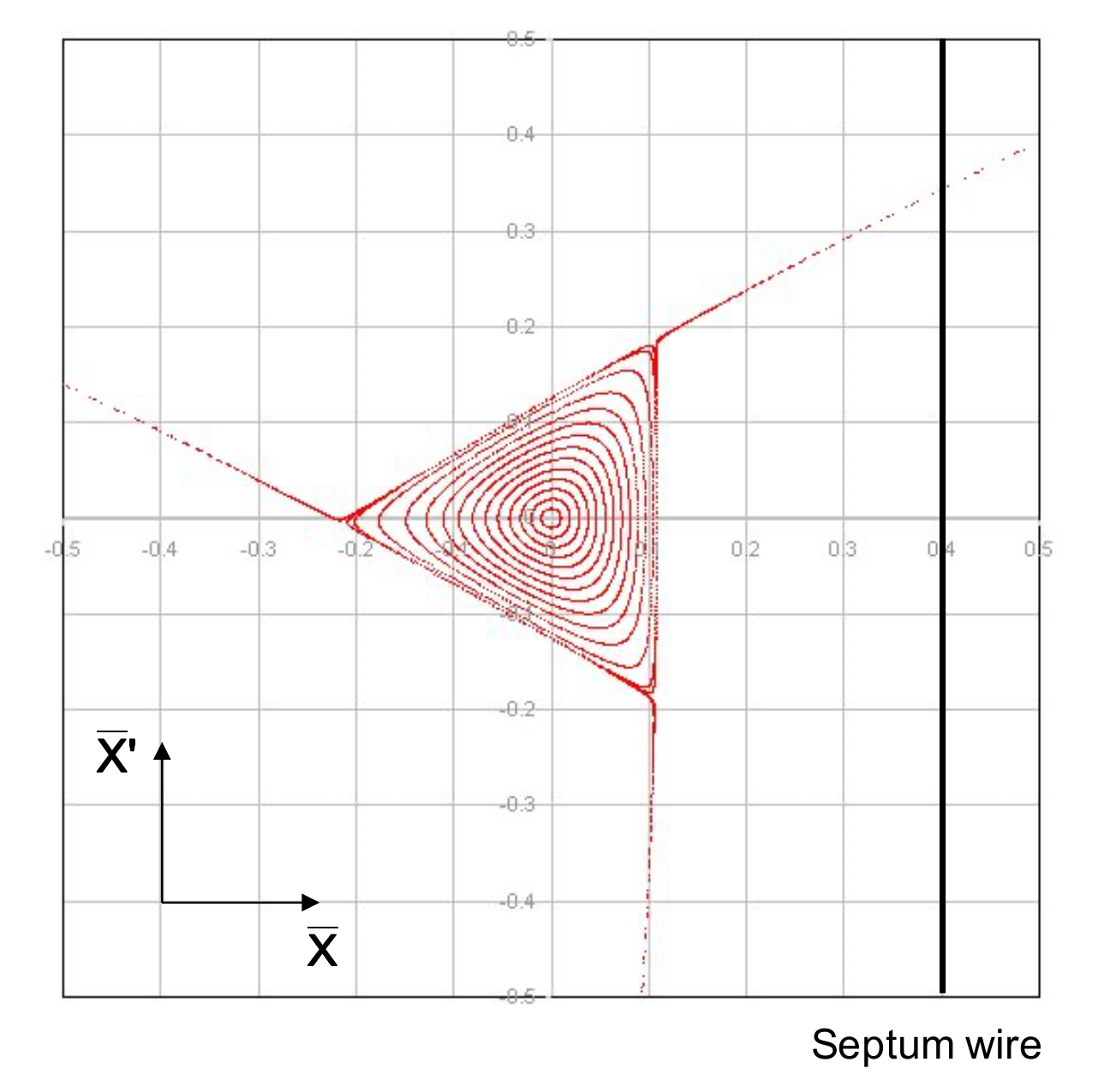}
e)\includegraphics[width=0.3\linewidth]{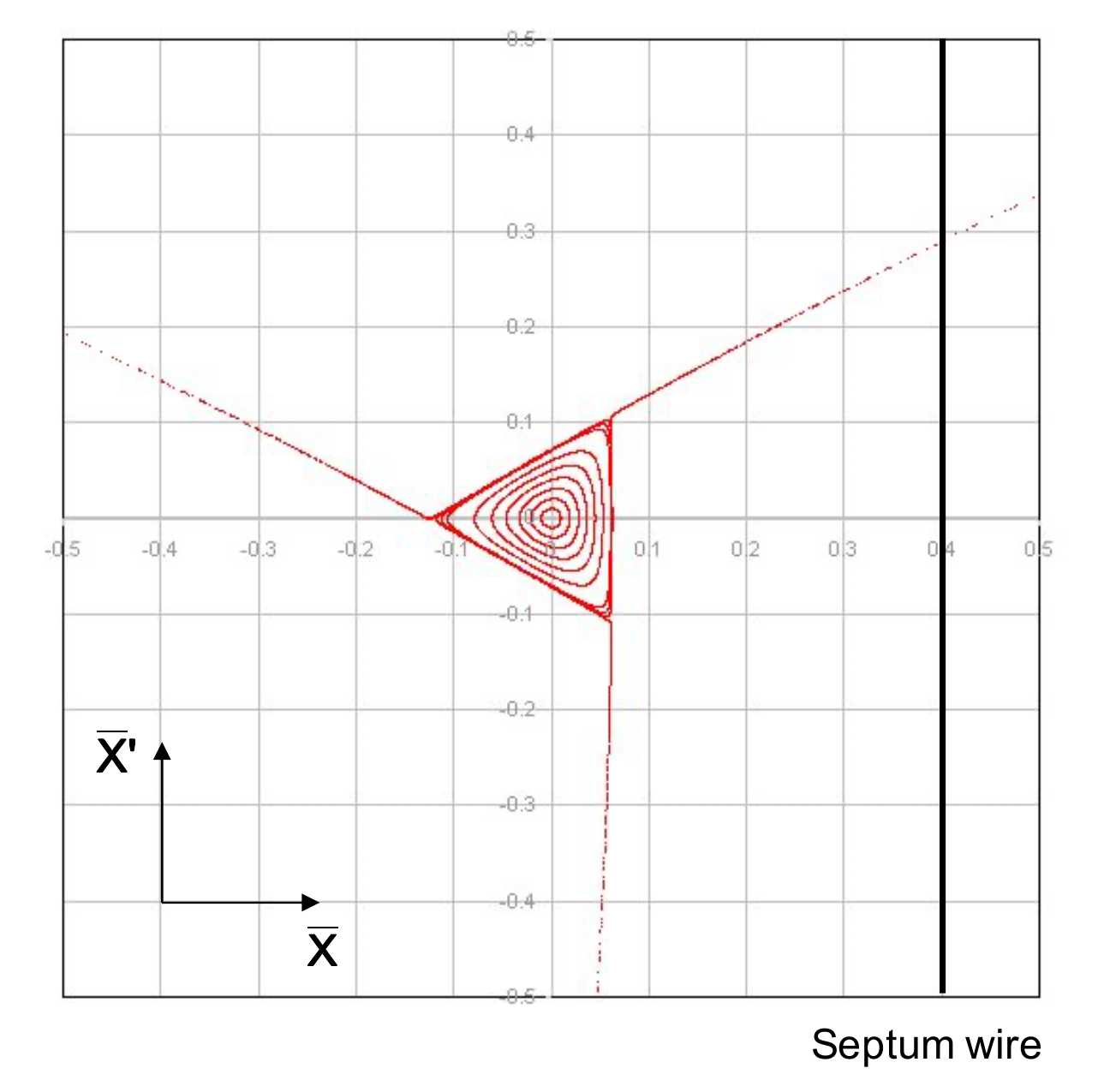}
f)\includegraphics[width=0.3\linewidth]{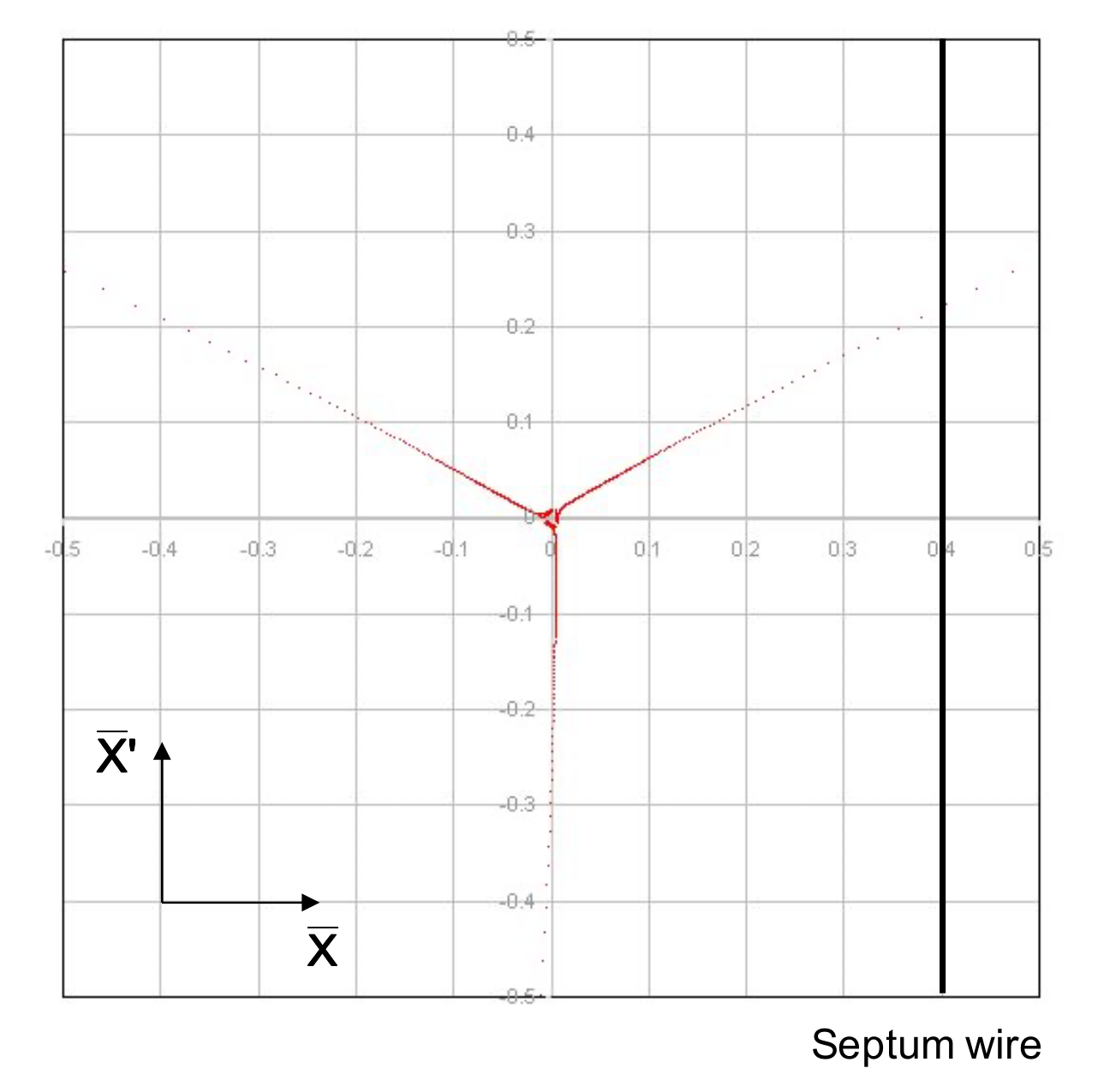}
\caption{Phase space evolution for resonant multi-turn extraction by a third-order resonance: 
a) $\Delta Q$ is large, there is no phase space distortion.
b) Sextupole magnets produce a triangular stable area in phase space,
with $\Delta Q$ decreasing, a phase space distortion appears first for largest amplitudes.
c) Largest amplitude particle trajectories are significantly distorted,
the locations of fixed points are noticeable at the extremities of phase space triangle.
d) $\Delta Q$ small enough that largest amplitude particle trajectories are unstable,
unstable particles follow separatrix branches as they increase in amplitude.
e) Stable area shrinks as $\Delta Q$  becomes smaller, separatrix position in phase space shifts,
the circulating beam intensity drops since particles are being continuously extracted.
f) As $\Delta Q$ approaches zero, the particles with very small amplitude are extracted.
}
\label{fig:slow-extraction}
\end{figure}

In the case of $3^\mathrm{rd}$-order resonances (see~\cite{bib:yp}),
sextupole fields distort the particles' circular normalised phase space trajectories.
The stable area is defined, delimited by the unstable Fixed Points, with
\begin{equation}
R_\mathit{fp}^{1/2} \propto \Delta Q \frac{1}{k_2}
\hspace*{2cm}\includegraphics[valign=c,width=0.31\linewidth]{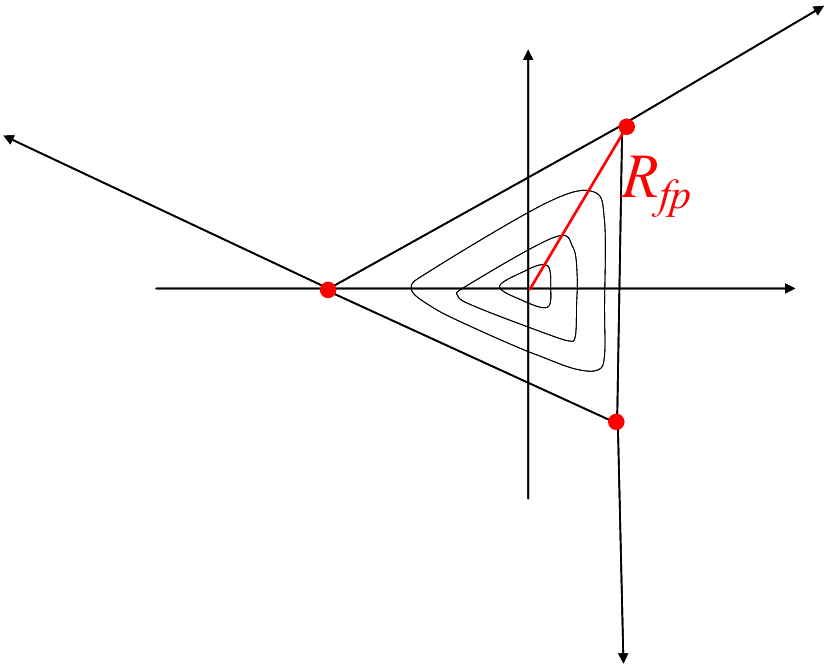}
\end{equation}
where $\Delta Q = Q - Q_r$, with $Q$ the betatron tune of the accelerator and $Q_r$ the
tune of the resonance, and $k_2$ the sextupole strength.
The sextupole magnets are arranged to produce a suitable phase space orientation of the stable triangle at 
the thin electrostatic septum.
The stable area can be reduced by increasing the sextupole strength $k_2$, or by scanning the machine tune $Q$
(and therefore the resonance) through the tune spread of the beam.
The slow extraction mechanism is illustrated in Fig.\ref{fig:slow-extraction}.

\section{\label{sec:linking}Linking accelerators}
\begin{figure}[!tb]
\centering\includegraphics[width=0.8\linewidth]{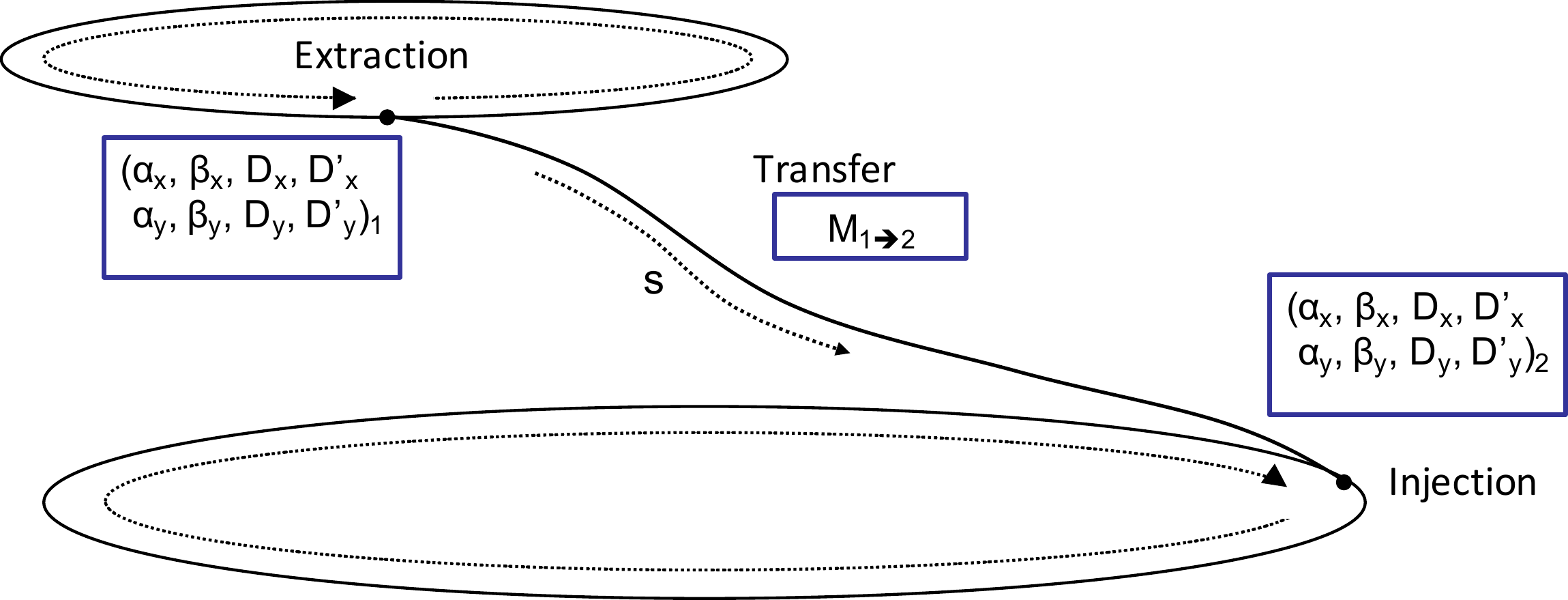}
\caption{Beam transfer from one accelerator to another. In addition to the proper geometry of the transfer line,
also the optics functions need to be matched in both transverse planes. }
\label{fig:linking}
\end{figure}

When beams need to be transported from the extraction of one machine to a fixed target or 
the injection into the next machine (see Fig.\ref{fig:linking}),
obviously the trajectory must be matched in all six geometric degrees of freedom (x,y,z,$\theta$,$\phi$,$\psi$).
 
But not only the geometry of the line is important, also the beam optics parameters of the injected beam 
have to be properly adapted to the beam optics of the receiving accelerator or the beam size at the target.
This implies that we need to “match” eight variables: $\alpha_x, \beta_x, D_x, D'_x$ and $\alpha_y, \beta_y, D_y, D'_y$.
Linking the optics is a complicated process due to the large number of parameters and often other constraints
from the geometry of the line or aperture restrictions.
%The parameters at start of line - given by the beam at this point - have to be propagated to
%matched parameters at the end of the line (injection to another machine, fixed target etc. )
The matching is achieved with a number of independently powered (so-called “matching”) quadrupoles.
Along the transfer line, maximum $\beta$ and dispersion values are imposed by magnetic apertures, and further
constraints can exist, like phase conditions for collimators, insertions for special equipment like stripping foils, etc.
So matching is done in practice with computer codes relying on a mixture of theory, experience,
intuition, and trial and error.

\begin{figure}[!tb]
\centering\includegraphics[width=0.8\linewidth]{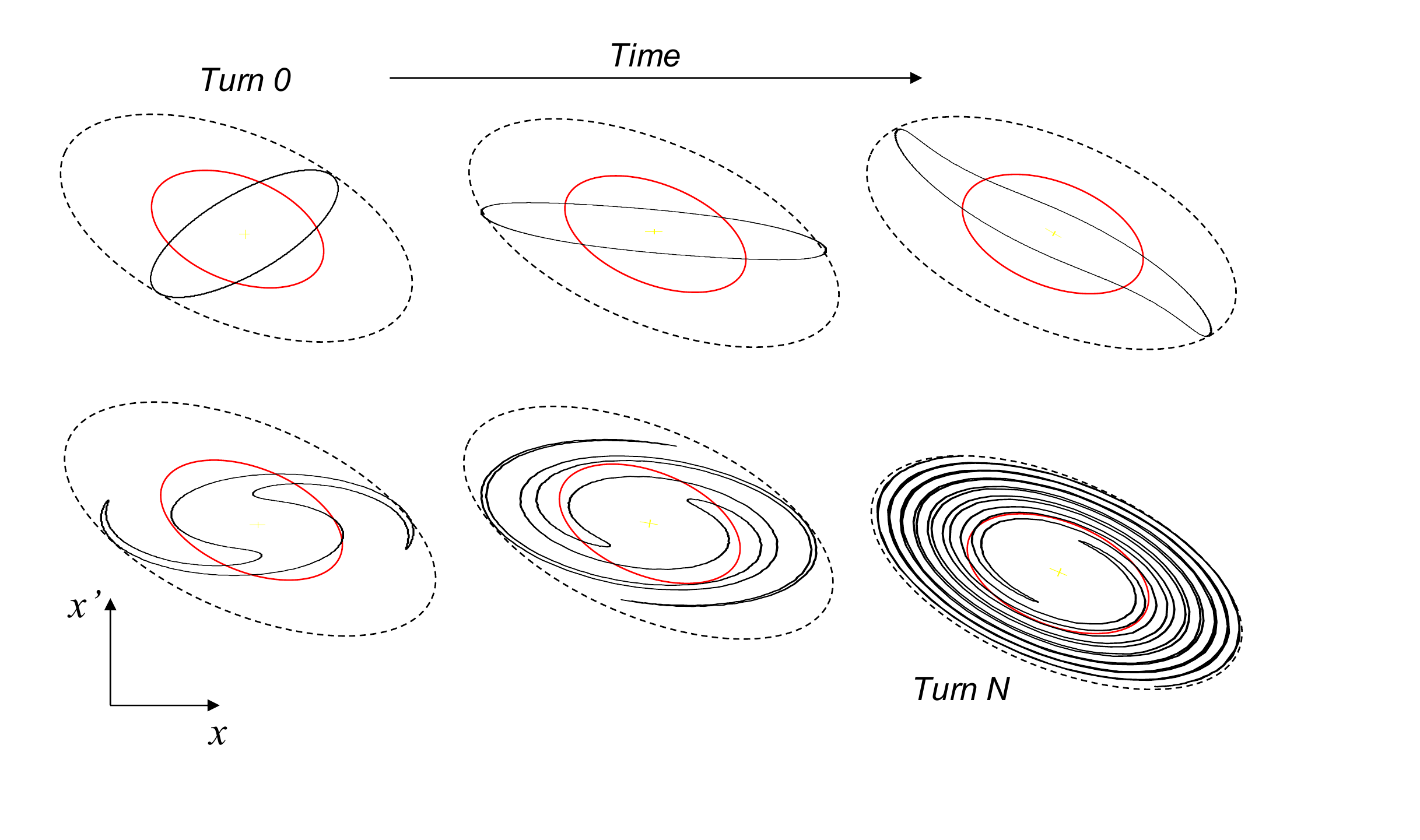}
\caption{Effect from optics mismatch where the phase space distribution of the incoming beam (in solid black)
does not correspond to the shape of the phase space ellipses of the receiving accelerator (in red).
Due to non-linear forces, the distribution will filament, leading to an effective emittance growth.  }
\label{fig:mismatch}
\end{figure}

When the optics parameters are not properly matched, i.e. the shape of the injected beam in phase space
does not correspond to the phase space contours of the closed optics solution, the different tunes for the
individual particles will lead to filamentation and emittance growth, as in the case of
a steering error for the injection (see Fig.~\ref{fig:mismatch}).
The same way, also an error in the dispersion functions and their derivatives will lead to an emittance
growth.

\section*{Acknowledgements}
I would like to thank my colleagues M.J.~Barnes, W.~Bartmann, J.~Borburgh, V.~Forte, M.~Fraser, 
B.~Goddard, V.~Kain and M.~Meddahi who have given this lecture at previous occasions of the CERN
Accelerator School course. The material presented here is almost entirely based on their lectures.

\end{document}